\documentclass[journal]{IEEEtran}
\usepackage{epstopdf}
\usepackage{subfigure}
\usepackage{multirow}
\usepackage{amsfonts}
\usepackage{mathrsfs}
\usepackage{amssymb}
\usepackage{epstopdf}
\usepackage{subfigure}
\usepackage[table]{xcolor}
\usepackage{color}
\usepackage{colortbl}
\usepackage{tabularx}
\usepackage{bm}
\usepackage{booktabs}
\definecolor{mypink}{rgb}{.99,.91,.95}

\usepackage{cite}
\usepackage{color}

\ifCLASSINFOpdf
 \usepackage[pdftex]{graphicx}
\else
\fi
\begin{document}
%
\title{Tensor Alignment Based Domain Adaptation for Hyperspectral Image Classification}
%
%
%
\author{Yao~Qin,~\IEEEmembership{Student Member,~IEEE},
       ~Lorenzo~Bruzzone,~\IEEEmembership{Fellow,~IEEE},
       {and}~Biao~Li
       
\thanks{Manuscript received September 4, 2018. (\emph{Corresponding author: Lorenzo Bruzzone}.)}
\thanks{Y. Qin is with the College of Electronic Science, National University of Defense Technology, Changsha 410073, China and Department of Information Engineering and Computer Science, University of Trento, 38122 Trento, Italy (e-mail: yao.qin@unitn.it).}
\thanks{L. Bruzzone is with Department of Information Engineering and Computer Science, University of Trento, 38122 Trento, Italy (e-mail: lorenzo.bruzzone@ing.unitn.it).}
\thanks{B. Li is with the College of Electronic Science, National University of Defense Technology, Changsha 410073, China.}
}

%
%

\markboth{Journal of \LaTeX\ Class Files,~Vol.~14, No.~8, August~2015}%
{Shell \MakeLowercase{\textit{et al.}}: Bare Demo of IEEEtran.cls for IEEE Journals}
%



\maketitle

\begin{abstract}
This paper presents a tensor alignment (TA) based domain adaptation method for hyperspectral image (HSI) classification. To be specific, HSIs in both domains are first segmented into superpixels and tensors of both domains are constructed to include neighboring samples from single superpixel. Then we consider the subspace invariance between two domains as projection matrices and original tensors are projected as core tensors with lower dimensions into the invariant tensor subspace by applying Tucker decomposition. To preserve geometric information in original tensors, we employ a manifold regularization term for  core tensors  into the decomposition progress. The projection matrices and core tensors are solved in an alternating optimization manner and the convergence of TA algorithm is analyzed. In addition, a post-processing strategy is defined via pure samples extraction for each superpixel to further improve classification performance.
Experimental results on four real HSIs demonstrate that the proposed method can achieve better performance compared with the state-of-the-art subspace learning methods when a limited amount of source labeled samples are available.
\end{abstract}


%
\IEEEpeerreviewmaketitle

\section{Introduction}
%
%
%
%
\IEEEPARstart{I}{n the} past decades, extensive research efforts have been spent on hyperspectral remote sensing since hyperspectral data contains detailed spectral information measured in contiguous bands of the electromagnetic spectrum \cite{Fauvel2013,Camps-Valls2014,Tuia2016a}. Due to the discriminative spectral information of such data, they have been used for a wide variety of applications, including agricultural monitoring \cite{Schneider2014}, mineral exploration \cite{Tiwari2011}, and \emph{etc}. One fundamental challenge in these applications is how to generate accurate land-cover maps.
Although supervised learning for hyperspectral image (HSI) classification has been extensively developed in the literature  (including random forest \cite{Ham2005}, support vector machine (SVM) \cite{Melgani2004}, laplacian SVM (LapSVM) \cite{Belkin2006,Melacci2011,Yang2015}, decision trees \cite{Delalieux2012} and support tensor machine (STM) \cite{Guo2016}), sufficient labeled training samples should be available to obtain satisfactory classification results. This would require extensive and expensive field data collection compaigns. 
Furthermore, with the advance of newly-developed spaceborne hyperspectral sensors, large numbers of HSIs are easily collected and it is not feasible to timely label samples of the hyperspectral images as reference for training.
Therefore, only limited labeled samples are available in most real applications of hyperspectral classification.  
According to the statistical theory in supervised learning, the data to be classified are expected to follow the same probability distribution function (\emph{PDF}) of training data. However, since the physical conditions (i.e. illumination, atmosphere, sensor parameters, and \emph{etc.}) can hardly be the same when collecting data, \emph{PDF}s of training and testing data tend to be different (but related) \cite{Bruzzone2010}. Then how to apply the labeled samples of original HSI to the related HSI is challenging in such cases. These problems can be addressed by adapting models trained on a limited number of source samples (\emph{source domain}) to new but related target samples (\emph{target domain}). The problem should be further studied for the development of hyperspectral applications.

According to the machine learning and pattern recognition literature, the problem of adapting model trained on a source domain to a target domain is referred to as \emph{transfer learning} or \emph{domain adaptation} (DA) \cite{Bruzzone2010}. The main idea of transfer learning is to adapt the knowledge learned in one task to a related but different task. An excellent review of transfer learning can be found in \cite{Patel2015,Pan2010}. In general, transfer learning is divided into four categories based on the properties of domains and tasks, i.e.  \emph{DA},  \emph{multi-task learning},  \emph{unsupervised transfer learning} and  \emph{self-taught learning}. In fact, DA has greater impact on practical applications. When applied to classification problems, DA aims to generate accurate classification results of target samples by utilizing the knowledge learned on the labeled source samples. According to \cite{Tuia2016a}, DA techniques for remote sensing applications can be roughly categorized as \emph{selection of invariant features}, \emph{adaptation of data distributions}, \emph{adaptation of classifier} and \emph{adaptation of classifier by active learning (AL)}.

In our case of HSI classification, we focus on the second category, i.e. \emph{adaptation of data distributions}, in which data distributions of both domains are made as similar as possible to keep the classifier unchanged. Despite the fact that several DA methods have been proposed for HSI classification, they treat HSIs as several single samples, which renders them incapable of reflecting and
preserving important spatial consistency of neighboring samples. 
In this paper, to exploit the spatial information in a natural and efficient way, \emph{tensorial processing} is utilized, which treats HSIs as three-dimensional (3D) tensors. Tensor arithmetic is a generalization of matrix and vector arithmetic, and is particularly well suited to represent multilinear relationships that neither vector nor matrix algebra can capture naturally \cite{Vasilescu2003,Lu2017}.
The power of tensorial processing for improving classification performance without DA has been proved in \cite{Zhong2015,Zhang2015,Xu2016,Feng2017,He2018}.
Similarly, when we apply tensorial processing to HSI in DA, multilinear relationships between neighboring samples in both HSIs are well captured and preserved, while conventional DA methods using vectorization deal with single samples. Tensor-based DA methods for visual application has demonstrated the efficacy and efficiency on the task of cross-domain visual recognition \cite{Lu2017,Koniusz2017}, whereas there are few published works on DA by using tensorial processing of HSIs. 

\begin{figure}[!t]
\setlength{\abovecaptionskip}{-5pt}
\setlength{\belowcaptionskip}{-20pt}
\centering
\includegraphics[width=0.5\textwidth]{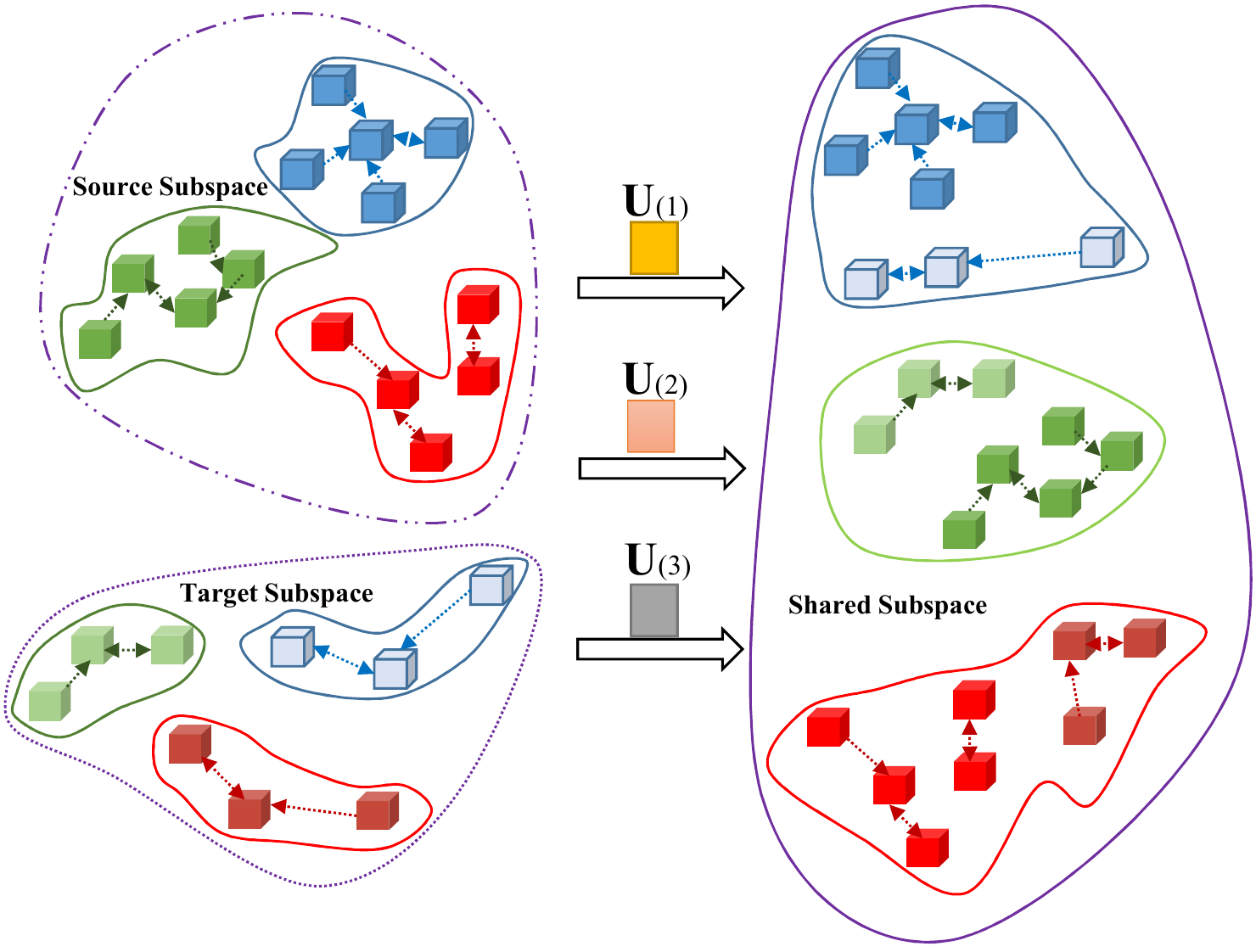}
\caption{Illustration of the manifold regularized tensor alignment method. There are 5 tensor objects for each class in the source domain, while 3 tensor objects for each class in the target domain. The shared subspace is obtained by utilizing 3 projection matrices $\{ \mathbf{U}^{(1)},\mathbf{U}^{(2)},\mathbf{U}^{(3)}\}$ with original geometry preserved. Each arrow represents the 1NN relationship between tensors. Best view in colors.}
\label{Flowchart}
\end{figure}

To be specific, we propose a tensor alignment (\emph{TA}) method for DA, which can be divided into two steps. First, the original HSI data cubes in both domains are divided into small superpixels and each central sample is represented as a 3D tensor consisting of samples in the same superpixel. In this way, each tensor is expected to include more samples from the same class. Since tensors are acted as ``\emph{basic elements}'' in the \emph{TA} method, we believe that high purity of tensors brings better adaptation performance. Second, taking into account the computational cost, we randomly select part of target tensors in the progress of \emph{TA} to identify the invariance subspace between the two domains as projection matrices. This is done on the source and selected target tensors, and the subspace shared by both domains is obtained by utilizing three projection matrices $\{ \mathbf{U}^{(1)},\mathbf{U}^{(2)},\mathbf{U}^{(3)}\}$ with original geometry preserved. The solution is addressed through the \emph{Tucker decomposition} \cite{Kolda2009} with orthogonal regularization on projection matrices and original tensors are represented by core tensors in the shared subspace. Fig. \ref{Flowchart} illustrates the manifold regularized \emph{TA} method with a 1-Nearest Neighbor (1NN) geometry preserved.

In addition to the \emph{TA} method, after generating classification map, a post-processing strategy based on \emph{pure  samples} extraction of each superpixel is employed to improve performances. The pure samples in superpixels have similar spectral features and likely belong to the same class. Therefore, if most pure samples in a superpixel are classified as $i$-th class, it is probable that the remaining pure samples belongs to the same class. 
Since samples in one superpixel may belong to two or even more classes and there are always classification errors in DA, the ratio of pure samples predicting as the same class might be reduced if we extract more pure samples.
Therefore, we extract the pure samples by fixing the ratio as 0.7. Specifically, final pure samples are extracted by first projecting samples in each superpixel to principal component axis and then including more samples in the middle range of the axis till the ratio reaches 0.7. In this way the consistency of classification results on pure samples is enforced.
To sum up, the main contributions of our work lie in the following two aspects:\\
\begin{tabularx}{0.5\textwidth}{lX}
$\! \!  \! \! \! \! \bullet \! \! \! \! \! $& {We propose a manifold regularized tensor alignment (\emph{TA}) for DA and develop the corresponding efficient iterative algorithm to find the solutions. Moreover, we analyze the convergence properties of the proposed algorithm and  its computational complexity as well.} \\
$\! \!  \! \! \! \! \bullet \! \! \! \! \! $& {We introduce a pure samples extraction strategy as post-processing to further improve the classification performance.} \\
\end{tabularx}
Comprehensive experiments on four publicly available benchmark HSIs have been conducted to demonstrate the effectiveness of the proposed algorithm.

The rest of the paper is organized as follows. Related works on adaptation of data distributions, tensorial processing of HSI and multilinear algebra are illustrated in Section II. The proposed methodology of \emph{TA} is presented in section III, while the pure samples extraction strategy for classification improvement is outlined in Section IV. Section V describes the  experimental datasets and setup. Results and discussions are presented in Section VI. Section VII summarizes the contributions of our research.

\section{Related Work}
This section briefly describes important studies related to the adaptation of data distributions, tensorial processing of hyperspectral data and basic concepts in multilinear algebra.
\subsection{Adaptation of Data Distributions}
Several methods for the adaptation of data distributions focus on subspace learning, where projected data from both domains are well aligned. Then, the same classifier (or regressor) is expected to be suitable for both domains. 
In \cite{Nielsen2009}, the data alignment is achieved through principal component analysis (PCA) or kernel PCA (KPCA).
In \cite{Fernando2013}, a PCA-based subspace alignment (SA) algorithm is proposed, where the source subspace is aligned as close as possible to the target subspace using a matrix transformation. 
In \cite{Lu2017}, features from convolutional neural network are treated as tensors and their invariant subspace is obtained through the Tucker decomposition. In \cite{Nielsen2007}, the authors align domains with canonical correlation analysis (CCA) and then perform change detection. The approach
is extended to a kernel and semisupervised version in \cite{Volpi2015}, where the authors perform change detection with
different sensors. In \cite{Samat2017}, the supervised multi-view canonical correlation analysis ensemble  is presented to address heterogeneous domain adaptation problems.

A few studies assume that data from both domains lie on the Grassmann manifold, where data alignment is conducted. In \cite{Gopalan2011}, the sampling geodesic flow (SGF) method is introduced and finite intermediate subspaces are sampled along the geodesic path connecting the source subspace and the target subspace. Geodesic flow kernel (GFK) method in \cite{Gong2012} models infinite subspaces in the way of incremental changes between both domains. Along this line, GFK support vector machine in \cite{Samat2016} shows the performance of GFK in nonlinear feature transfer tasks. A GFK-based hierarchical subspace learning strategy for DA is proposed in \cite{Banerjee2017}, and an iterative coclustering technique applied to the subspace obtained by GFK is proposed in \cite{Banerjee2016}.

Other studies hold the view that the subspace of both domains can be low-rank reconstructed or clustered. The reconstruction matrix is enforced to be low-rank and a sparse matrix is used to represent noise and outliers.
In \cite{Jhuo2012}, a robust domain adaptation low-rank reconstruction (RDALRR) method is proposed, where a transformed intermediate representation of the samples in the source domain is linearly reconstructed by the target samples.
In \cite{Shao2014}, the low-rank transfer subspace learning (LTSL) method is proposed where transformations are applied for both domains to resolve disadvantages of RDALRR. In \cite{Xu2016a}, a low-rank and sparse representation (LRSR) method is presented by additionally enforcing the reconstruction matrix to be sparse. To obtain better results of reconstruction matrix, structured domain adaptation (SDA) in \cite{Li2017a} utilizes block-diagonal matrix to guide iteratively the computation. Different from the above methods, latent sparse domain transfer (LSDT) in \cite{Zhang2016a} is inspired by subspace clustering, while the low-rank reconstruction and instance weighting label propagation algorithm in \cite{Shi2015} attempts to find new representations for the samples in different classes from the source domain by multiple linear transformations. 

Other methods focus on feature extraction strategy by minimizing predefined distance measures, e.g. \emph{Maximum Mean Discrepancy} (MMD) or Bregman divergence. In \cite{Pan2011}, transfer component analysis (TCA)  tries to learn some
transfer components across domains in a Reproducing Kernel Hilbert Space (RKHS) using MMD. It is then applied to remote sensing images in \cite{Matasci2015}. TCA is further improved by joint domain adaptation (JDA), where both the marginal distributions and conditional distributions are adapted in a dimensionality reduction procedure  \cite{Long2013}. Furthermore,  transfer joint matching (TJM) aims to reduce the domain difference by jointly matching the features and reweighting the instances across domains \cite{Long2014}. Recently, joint geometrical and statistical alignment (JGSA) is presented by reducing the MMD and forcing both projection matrices to be close \cite{Zhang2017}. 
In \cite{Sun2016a} and \cite{Sun2016}, the authors transfer category models trained on landscape views to aerial views for high-resolution remote sensing images by reducing MMD. 

Different from the above category for feature extraction, several studies employ manifold learning to preserve the original geometry. In \cite{Tuia2014}, both domains are matched through manifold alignment while preserving label (dis)similarities and the geometric structures of the single manifolds. The algorithm is extended to a kernelized version in \cite{Tuia2016}. Spatial information of HSI data is taken into account for manifold alignment in \cite{Yang2016}. In \cite{Yang2016a}, both local and global geometric characteristics of both domains are preserved and bridging pairs are extracted for alignment.
In addition to manifold learning, the manifold regularized domain adaptation (MRDA) method integrates spatial information and the overall mean coincidence method to improve prediction accuracy \cite{Luo2018}.
Beyond classical subspace learning, manifold assumption and feature extraction methods, several other approaches are proposed in the literature, such as class centroid alignment \cite{Zhu2016}, histogram matching \cite{Inamdar2008} and graph matching \cite{Tuia2013}.

\subsection{Tensorial Processing of Hyperspectral Data}
A few published works of tensor-based methods have been applied to HSI processing to fully exploit both spatial and spectral information. The texture features of HSI at different scales, frequencies and orientations are successfully extracted by the 3D discrete wavelet transform (3D-DWT) \cite{Qian2013}.  The gray level  co-occurrence  is  extended  to  its  3D  version in \cite{Tsai2013} to improve classification  performance.  Tensor  discriminative  locality alignment (TDLA) algorithm optimizes the  discriminative  local  information for feature extraction  \cite{Zhang2013}, while local tensor discriminant analysis (LTDA) technique is employed in \cite{Zhong2015a} for spectral-spatial feature extraction. The high-order structure of HSI along all dimensions is fully exploited by superpixel tensor sparse coding to better understand the  data in \cite{Feng2017a}.  
Moreover, several conventional 2D methods are extended to the 3D for HSI processing, such as the 3D extension of empirical mode decomposition in \cite{He2017,He2016}.  The modified tensor locality preserving projection (MTLPP) algorithm is presented for HSI dimensionality reduction and classification in \cite{Deng2018}.

\subsection{Notations and basics of Multilinear Algebra}
A tensor is a multi-dimensional array that generalizes matrix representation. Vectors and matrices are first and second order tensors, respectively. In this paper, we use lower case letters (e.g. $x$), boldface lowercase letters (e.g. $\mathbf{x}$) and boldface capital
letters (e.g.  $\mathbf{X}$) to denote scalars, vectors and matrices, respectively.
Tensors of order 3 or higher will be denoted by
boldface Euler script calligraphic letters (e.g.  ${\mathcal{X}}$). The operations of Kronecker product, Frobenius norm, vectorization and product are denoted by $\otimes$, $||\cdot||_{\mathrm{F}}$, $\mathrm{vec}(\cdot)$ and $\prod$, respectively. The $\mathrm{Tr}(\cdot)$ denotes the trace of a matrix.

A \emph{M}-th order tensor is denoted by $\mathcal{X} \in \mathbb{R}^{I_{1}\times\cdots \times I_{M} }$. 
The corresponding $k$-mode matricization of the tensor $\mathcal{X}$, denoted by  $\mathbf{X}_{(k)}$, unfolds the tensor $\mathcal{X}$ with respect to mode $k$.
The operation between tensor $\mathcal{X}$  and a matrix $\mathbf{U}^{(k)} \in \mathbb{R}^{J_{k} \times I_{k}}$ is the $k$-mode product denoted by $\mathcal{X} \times_{(k)}\mathbf{U}^{(k)}  $, which is a tensor of size ${I_{1} \times \cdots \times I_{k-1} \times J_{k} \times I_{k+1} \times \cdots I_{M} }$.
Briefly, the notion for the product of tensor with a  set of projection matrices \{$\mathbf{U}^{(k)} \}_{k=1}^{M}$ excluding the $l$-mode as:
\begin{equation}
\mathcal{X}\bar{\times}_{(l)}\mathbf{U}^{(l)} =\mathcal{X} \prod_{k\neq l }\times_{(k)}\mathbf{U}^{(k)}
\end{equation}
Tucker decomposition is one of the most well-known decomposition models for tensor analysis. It decomposes a $M$ mode tensor ${\mathcal{X}}$  into a core tensor ${\mathcal{G}}$ multiplied by a set projection matrices \{$\mathbf{U}^{(k)} \}_{k=1}^{M}$ with the objective fuction defined as follows:
\begin{equation}
\min_{\mathcal{G},\mathfrak{U}} \arrowvert\arrowvert \mathcal{X}-\mathcal{G} \prod_{k=1}^{M}\times_{(k)}\mathbf{U}^{(k)} \arrowvert\arrowvert _{\mathrm{F}}
\end{equation}
where $\mathcal{G} \in \mathbb{R}^{J_{1}\times\cdots \times J_{M} }$ and $\mathfrak{U}$ represents \{$\mathbf{U}^{(k)} \}_{k=1}^{M}$ with $\mathbf{U}^{(k)} \in \mathbb{R}^{J_{k} \times I_{k}}$. For simiplicity, we denote $\mathcal{G} \prod_{k=1}^{M}\times_{(k)}\mathbf{U}^{(k)}$ as $\lbrack\lbrack \mathcal{G};\mathfrak{U}\rbrack\rbrack$. 

By applying the $l$-mode unfolding, Eq. (2) can alternatively be written as 
\begin{equation}
\min_{\mathcal{G},\mathfrak{U}} \arrowvert\arrowvert \mathbf{X}_{(l)}-\mathbf{U}^{(l)}\mathbf{G}_{(l)}(\mathbf{U}^{(-l)})^{\mathrm{T}} \arrowvert\arrowvert _{\mathrm{F}}
\end{equation}
where $\mathbf{G}_{(l)}$ denotes the $l$-mode unfolding of $\mathcal{G}$, and $\mathbf{U}^{(-l)}$ denotes  $\prod_{k\neq l }\otimes\mathbf{U}^{(k)}$.
The vectorization of (3)  can be formulated as
\begin{equation}
\min_{\mathcal{G},\mathfrak{U}} \arrowvert\arrowvert \mathrm{vec}(\mathbf{X}_{(l)})-(\mathbf{U}^{(-l)}\otimes\mathbf{U}^{(l)})\mathrm{vec}(\mathbf{G}_{(l)})\arrowvert\arrowvert _{\mathrm{F}}  
\end{equation}
Note that regularizations of $\mathcal{G}$ and $\mathfrak{U}$ are ignored in above equations. 

\section{Proposed Tensor Alignment Approach}
\subsection{Problem Definition}
Assume that we have $n_{s}$ tensor samples $\{\mathcal{X}_{s}^{i}\}_{i=1}^{n_{s}}$ of mode $M$ in the source domain, where $\mathcal{X}_{s}^{i}\in\mathbb{R}^{I_{1}\times\cdots \times I_{M}}$. Similarly, tensors in the target domain are denoted as $\{\mathcal{X}_{t}^{j}\}_{j=1}^{n_{t}}$. In this paper, we consider only homogeneous DA problem, thus we assume that $\mathcal{X}_{t}^{j} \in \mathbb{R}^{I_{1}\times\cdots\times I_{M}}$. 
In the context of DA, we follow the idea to represent the subspace invariance between two domains as projection matrices $\mathfrak{U}=\{ \mathbf{U}^{l} \}_{l=1}^{M}$, where $\mathbf{U}^{l}\in\mathbb{R}^{J_{l}\times I_{l}}$.
Intuitively, we propose to conduct subspace learning on the tensor samples in both domains with manifold regularization. 
Fig. \ref{Flowchart} shows that the shared subspace is obtained by utilizing 3 projection matrices $\{ \mathbf{U}^{(1)},\mathbf{U}^{(2)},\mathbf{U}^{(3)}\}$. By performing Tucker decomposition simultaneously, the tensor samples in both domains are represented by the corresponding core tensors $\mathcal{G}_{\mathcal{S}}=\{\mathcal{G}_{\mathcal{S}}^{i}\}_{i=1}^{n_s}$ and $\mathcal{G}_{\mathcal{T}}=\{\mathcal{G}_{\mathcal{T}}^{i}\}_{i=1}^{n_T}$ with smaller dimensions in the shared subspace. The geometrical information should be preserved as much as possible via forcing manifold regularization during subspace learning.
In the next subsection, we will introduce how to construct HSI tensors and perform tensor subspace alignment.
\begin{figure}[!t]
\setlength{\abovecaptionskip}{-5pt}
\setlength{\belowcaptionskip}{-20pt}
\centering
\includegraphics[width=0.5\textwidth]{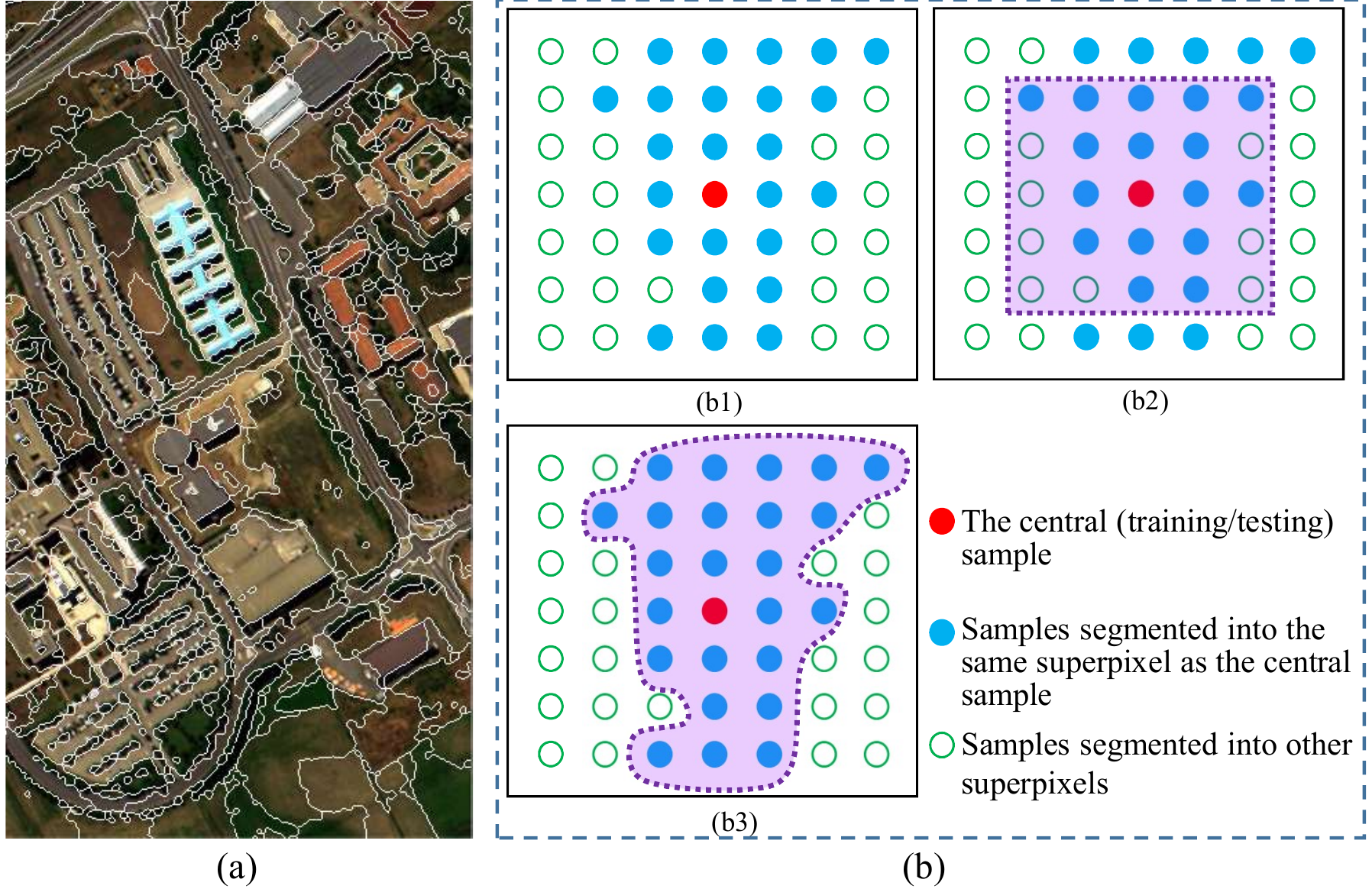}
\caption{Illustration of (a) superpixel segmentation of Pavia University data and (b) how to determine the $5 \times 5$ neighbors of the central (training/testing) sample. (b1) A $7 \times 7$ patch surrounded by the central sample. (b2) Conventional methods take the $5 \times 5$ patch as neighbors (see the purple area). (b3) Strategy used in our method.}
\label{Construction}
\end{figure}
\subsection{Tensors Construction}
DA is achieved by tensor alignment, where HSI tensors are regarded as ``\emph{basic elements}''. Therefore, HSI tensors are expected to be as pure as possible. An ideal tensor should consist of samples belonging to the same class. However, spatial square patches  centered at the training (or testing) samples can contain samples from different classes when extracted at the edge between different classes. To obtain pure tensors, superpixel segmentation \cite{Achanta2012} is performed on the first three  principal components of the HSI. Fig. \ref{Construction}(a) shows the segmentation result of Pavia University data as an illustrative example. Then, samples surrounding the central sample in the same superpixel are utilized to form the tensor [see Fig. \ref{Construction}(b)]. This strategy can preserve the spectral-spatial structure of the HSI cube and takes into account the dissimilarity of various neighborhoods, particularly at the edges of different classes.

\subsection{Method Formulation}
After the construction of tensors, two weight matrices for both domains are computed to enforce manifold regularization. 
We compute the source weight matrix $\mathbf{W}_{\mathcal{S}}$ in a supervised manner with labels of source tensors, while 10 nearest neighboring samples are searched via spectral angle measure (SAM) for target weight matrix $\mathbf{W}_{\mathcal{T}}$.  
Then, the binary distance is employed for weight matrix construction. To reduce computation cost, dimensionality reduction for all tensors are conducted via multilinear PCA (MPCA). 
Given tensor samples and weight matrices for both domains, the final optimization problem is then defined as:
\begin{eqnarray}
 \!  \min_{\mathcal{G}_{\mathcal{S}},\mathcal{G}_{\mathcal{T}},\mathfrak{U}} \sum_{i=1}^{n_\mathcal{S}}\!||\mathcal{X}_{\mathcal{S}}^{i}-[[\mathcal{G}_{\mathcal{S}}^{i};\mathfrak{U}]]||_{\mathrm{F}}^{2}+\sum_{j=1}^{n_\mathcal{T}}\!||\mathcal{X}_{\mathcal{T}}^{j}-[[\mathcal{G}_{\mathcal{T}}^{j};\mathfrak{U}]]||_{\mathrm{F}}^{2}  \nonumber \\
 \!+\!{\lambda} (\sum_{i=1}^{n_\mathcal{S}} \sum_{j=1}^{n_\mathcal{S}}\!||\mathcal{G}_{\mathcal{S}}^{i}-\mathcal{G}_{\mathcal{S}}^{j}||_{\mathrm{F}}^{2}w_{ij}^{\mathcal{S}}\!+\!\sum_{i=1}^{n_\mathcal{T}} \sum_{j=1}^{n_\mathcal{T}}\!||\mathcal{G}_{\mathcal{T}}^{i}-\mathcal{G}_{\mathcal{T}}^{j}||_{\mathrm{F}}^{2}w_{ij}^{\mathcal{T}}) \nonumber \\
\! \! \! \! \mathrm{s.t.} \ \ \ {\mathbf{U}}^{(l)\mathrm{T}}\mathbf{U}^{(l)}= \mathbf{I}, 1 \le l \le M\ \ \ \ \  \ \ \ \ \ \ \ \  \ \ \ \ \  \ \ \ \ \ \ \ 
\end{eqnarray}
where $\lambda$ is a nonnegative parameter for controlling the importance of manifold regularization, $w_{ij}^{\mathcal{S}}$ and $w_{ij}^{\mathcal{T}}$ are weight between the $i$-th and $j$-th tensor in $\mathbf{W}_{\mathcal{S}}$ and $\mathbf{W}_{\mathcal{T}}$, respectively. When $\lambda$ is small,
the objective function depends mainly on the minimization of reconstruction errors for all tensor objects. When it is large, the objective function depends mainly on the preservation of tensor geometry information.

\subsection{Optimization}
The problem in (5) can be solved by alternatively updating $\mathfrak{U}$ and the cores  $\mathcal{G}_{\mathcal{S}}$ and  $\mathcal{G}_{\mathcal{T}}$  until the objective function converges.
Herein, by applying the $k$-mode unfolding and according to (3), we obtain the following problem
\begin{eqnarray}
\min_{\mathcal{G}_{\mathcal{S}},\mathcal{G}_{\mathcal{T}},\mathfrak{U}} \sum_{\mathscr{D}\in\{\mathcal{S},{\mathcal{T}}\}}\sum_{i=1}^{n_\mathscr{D}}\!||\mathbf{X}_{\mathscr{D}(k)}^{i}\!-\!\mathbf{U}^{(k)}\mathbf{G}_{\mathscr{D}(k)}^{i}(\mathbf{U}^{(-k)})^{\mathrm{T}}\!||_{\mathrm{F}}^{2} \  \ \nonumber \\
+{\lambda}\sum_{\mathscr{D}=\mathcal{S}}^{\mathcal{T}}\sum_{i=1}^{n_\mathscr{D}} \sum_{j=1}^{n_\mathscr{D}}||\mathbf{G}_{\mathscr{D}(k)}^{i}-\mathbf{G}_{\mathscr{D}(k)}^{j}||_{\mathrm{F}}^{2}w_{ij}^{\mathscr{D}}  \ \ \ \  \nonumber \\
\! \! \! \! \mathrm{s.t.} \ \ \ {\mathbf{U}}^{(l)\mathrm{T}}\mathbf{U}^{(l)}= \mathbf{I},1 \le l \le M\ \ \ \ \  \ \ \ \ \ \ \ \  \ \ \ \ \  \ \ 
\end{eqnarray}
where $\mathscr{D}$ is introduced to denote source ($\mathcal{S}$) and target ($\mathcal{T}$) domain for simplicity.
\subsubsection{Updating $\mathcal{G}_{\mathcal{S}}$ and  $\mathcal{G}_{\mathcal{T}}$}
When $\mathfrak{U}$ is fixed,  by applying vectorization shown in (4), the problem is formulated as 
\begin{eqnarray}
\min_{\mathcal{G}_{\mathcal{S}},\mathcal{G}_{\mathcal{T}}}\sum_{\mathscr{D}\in\{\mathcal{S},{\mathcal{T}}\}}\sum_{i=1}^{n_\mathscr{D}}\!||\mathrm{vec}(\mathbf{X}_{\mathscr{D}(k)}^{i})\!-\!\mathbf{Z}^{(k)}\mathrm{vec}(\mathbf{G}_{\mathscr{D}(k)}^{i})||_{\mathrm{F}}^{2} \  \ \nonumber \\
+{\lambda} \sum_{\mathscr{D}\in\{\mathcal{S},{\mathcal{T}}\}}\sum_{i=1}^{n_\mathscr{D}} \sum_{j=1}^{n_\mathscr{D}}||\mathrm{vec}(\mathbf{G}_{\mathscr{D}(k)}^{i})-\mathrm{vec}(\mathbf{G}_{\mathscr{D}(k)}^{j})||_{\mathrm{F}}^{2}w_{ij}^{\mathscr{D}}  
\end{eqnarray}
where $\mathbf{Z}^{(k)}=\mathbf{U}^{(-k)}\otimes\mathbf{U}^{(k)}$.
The matrix form of the above equation can be written as
\begin{eqnarray}
\min_{\mathcal{G}_{\mathcal{S}},\mathcal{G}_{\mathcal{T}}} \sum_{\mathscr{D}\in\{\mathcal{S},{\mathcal{T}}\}}||\mathbf{X}_{\mathscr{D}(k)}^{v}-\mathbf{Z}^{(k)}\mathbf{G}_{\mathscr{D}(k)}^{v}\!||_{\mathrm{F}}^{2} \  \ \nonumber \\
+{\lambda}\sum_{\mathscr{D}\in\{\mathcal{S},{\mathcal{T}}\}}\mathrm{Tr}(\mathbf{G}_{\mathscr{D}(k)}^{v}\mathbf{L}^{\mathscr{D}}(\mathbf{G}_{\mathscr{D}(k)}^{v})^{\mathrm{T}}) 
\end{eqnarray}
where $\mathbf{X}_{\mathscr{D}(k)}^{v}$ and $\mathbf{G}_{\mathscr{D}(k)}^{v}$ are matrices in which the $i$-th columns are $\mathrm{vec}(\mathbf{X}_{\mathscr{D}(k)}^{i})$ and $\mathrm{vec}(\mathbf{G}_{\mathscr{D}(k)}^{i})$, respectively. 
The $\mathbf{L}^{\mathscr{D}}=\mathbf{D}^{\mathscr{D}}-\mathbf{W}^{\mathscr{D}}$ denotes the Laplacian matrix, $\mathbf{D}^\mathscr{D}=\mathrm{diag}(d_{1}^{\mathscr{D}},...,d_{n}^{\mathscr{D}})$ and $d_{i}^{\mathscr{D}}=\sum_{j=1}^{n_{\mathscr{D}}}\mathbf{W}_{ij}^{\mathscr{D}}$.
Formally, we transform the problem above into the following optimization formulation
\begin{equation}
\min_{\mathbf{G}} ||\mathbf{X}-\mathbf{ZG}||_{\mathrm{F}}^{2}+\lambda\mathrm{Tr}(\mathbf{GLG}^{\mathrm{T}})
\end{equation}
For simplicity and better illustration, $\mathbf{X}\in \mathbb{R}^{n_x \times n_D}$, $\mathbf{Z}\in \mathbb{R}^{n_x \times n_g}$ and $\mathbf{L}\in \mathbb{R}^{n_D \times n_D}$. 
Let 
 $ \mathbf{\Lambda \Sigma V}^{\mathrm{T}}$  be the singular value decomposition of $\mathbf{Z}$ and denote $\mathbf{V}^{\mathrm{T}}\mathbf{G}$ as $\mathbf{Y}$. Note that no information of $\mathbf{G}$ is lost in the transformation because $\mathbf{V}$ is invertible. Then we have
\begin{equation}
\min_{\mathbf{Y}} ||\mathbf{X}-\mathbf{ \mathbf{\Lambda  \Sigma Y}}||_{\mathrm{F}}^{2}+\lambda\mathrm{Tr}(\mathbf{VYLY}^{\mathrm{T}}\mathbf{V}^{\mathrm{T}})
\end{equation}
where $\mathbf{Y}\in\mathbb{R}^{n_g\times n_x}$, $\mathbf{\Lambda\Lambda}^{\mathrm{T}}=\mathbf{I}_{n_x \times n_x}$ and $\mathbf{VV}^{\mathrm{T}}=\mathbf{I}_{n_g \times n_g}$. Based on the properties of trace and F-norm, we reformulate it as 
\begin{equation}
\min_{\mathbf{Y}} ||\mathbf{M}-\mathbf{ \mathbf{  \Sigma Y}}||_{\mathrm{F}}^{2}+\lambda\mathrm{Tr}(\mathbf{YLY}^{\mathrm{T}})
\end{equation}
where $\mathbf{M}=\mathbf{\Lambda}^{\mathrm{T}}\mathbf{X}$. We denote the $i$-th row of matrix $\mathbf{M}$ as $\mathbf{M}_{i,:}$. Then the problem above can be rewritten as
\begin{equation}
\min_{\mathbf{Y}} \sum_{i=1}^{n_g}||\mathbf{M}_{i,:}-\mathbf{ \Sigma}_{ii}\mathbf{ Y}_{i,:}||_{\mathrm{F}}^{2}+ \lambda\sum_{i=1}^{n_g}\mathbf{Y}_{i,:}\mathbf{L}\mathbf{Y}_{i,:}^{\mathrm{T}}
\end{equation}
When only considering  $\mathbf{Y}_{i,:}$, we have
\begin{equation}
\min_{\mathbf{Y}_{i,:}}  \mathbf{Y}_{i,:}\mathbf{Q}\mathbf{Y}_{i,:}^{\mathrm{T}}-2\mathbf{ \Sigma}_{ii} \mathbf{ Y}_{i,:} \mathbf{M}_{i,:}^{\mathrm{T}} 
\end{equation}
where $\mathbf{Q}=\lambda\mathbf{L}+\mathbf{ \Sigma}_{ii}^{2}\mathbf{I}_{n_D \times n_D}$ is a positive definite matrix.
This is an unconstrained quadratic programming optimization of $\mathbf{Y}_{i,:}$ and can be easy solved by setting the derivation to zero. 
The optimal $\mathbf{Y}^{*}$ can be obtained by updating all rows and optimal $\mathbf{G}^{*}$ is given as $\mathbf{VY}^{*}$.
When both $\mathbf{G}_{\mathscr{D}(k)}^{v}$ are updated, the $\mathcal{G}_{\mathscr{D}}$ can be obtained by applying tensorization.

\subsubsection{Updating  $\mathfrak{U}$}
When the cores  $\mathcal{G}_{\mathcal{S}}$ and  $\mathcal{G}_{\mathcal{T}}$ are fixed, 
we first write the problem of (5) as
\begin{eqnarray}
\min_{\mathfrak{U}}||\mathcal{X}-\mathcal{G}\prod_{l=1}^{M}\mathbf{U}^{(l)}||_{\mathrm{F}}^{2}, {\mathbf{U}}^{(l)\mathrm{T}}\mathbf{U}^{(l)}= \mathbf{I}, 1 \le l \le M
\end{eqnarray}
where $\!\!\mathcal{X}\!\!\in\!\!\mathbb{R}^{I_1 \times \cdots \times I_M \times (n_{\mathcal{S}}+n_{\mathcal{T}})}$ and $\mathcal{G}\!\!\in\!\!\mathbb{R}^{J_1 \times \cdots \times J_M \times (n_{\mathcal{S}}+n_{\mathcal{T}})}$ denote the concatenation of sample tensors $\mathcal{X}_{\mathscr{D}}$ and core tensors $\mathcal{G}_{\mathscr{D}}$  in each domain, respectively.
Similar to most tensor decomposition algorithms, the solution for $\mathbf{U}^{(l)} (1 \le l \le M)$ is obtained by updating one with others fixed. 
By using the $k$-mode unfolding, the problem is derived as the following constrained objective function:
\begin{eqnarray}
\min_{\mathbf{U}^{(k)}} ||\mathbf{X}_{(k)}\!-\!\mathbf{U}^{(k)}\mathbf{G}(\mathbf{U}^{(-k)})^{\mathrm{T}}\!||_{\mathrm{F}}^{2},{\mathbf{U}}^{(k)\mathrm{T}}\mathbf{U}^{(k)}= \mathbf{I}
\end{eqnarray}
which can be effectively solved by utilizing  singular value decomposition (SVD) of $\mathbf{X}_{(k)}\mathbf{U}^{(-k)}\mathbf{G}^{\mathrm{T}}$. Please refer to the Appendix for the proof.
For an efficient computation, $\mathbf{G}(\mathbf{U}^{(-k)})^{\mathrm{T}}$ is solved in the implementation as follows:
\begin{equation}
\mathbf{G}(\mathbf{U}^{(-k)})^{\mathrm{T}}=[\mathcal{G}\prod_{l\ne k} \times_{(l)} \mathbf{U}^{(l)}]_{(k)}
\end{equation}
Based on the derived solutions of projection matrices and core tensors, the proposed method is summarized in Algorithm 1. Since the objective function in (5) is non-convex on projection matrices  $\mathfrak{U}$ and core tensors  $\mathcal{G}_{\mathscr{D}}$, we initialize projection matrices $\mathfrak{U}$ to obtain a stationary solution by solving a conventional Tucker decomposition problem.

\begin{displaymath}
\setlength{\arrayrulewidth}{0.4mm}
\begin{tabular}{p{8.5cm}}
\hline
$\! \mathbf{Algorithm \ 1}$ Tensor Alignment \\
\hline
$\mathbf{Input}$: tensor set $\mathcal{X}_{\mathscr{D}}$ in both  domains, regularization parameter $\lambda$ and dim of cores $\{ J_{1},\dots,J_{M} \}$ \\     
$\mathbf{Output}$:  core tensor set $\mathcal{G}_{\mathscr{D}}$ in both domains, projection matrices $\mathfrak{U}$ \\
1: Compute two graph matrice $\mathbf{W}_{\mathscr{D}}$;\\
2: Initialize $\mathfrak{U}$ using Tucker decomposition;\\
3: $\mathbf{While}$ optimization in (5) does not converge $\mathbf{do}$\\
4: \ \ \ \ \ update $\mathcal{G}_{\mathscr{D}}$ by solving (13); \\
5: \ \ \ \ \ update $\mathfrak{U}$ by alternatively solving (15); \\
6: $\mathbf{Check}$ the convergence of (5);\\
7: $\mathbf{Return}$ $\mathcal{G}_{\mathscr{D}}$ and $\mathfrak{U}$.\\
\hline
\end{tabular}
\end{displaymath}

\subsection{Convergence and Computational Complexity Analysis} 
Formally, the objective function of the optimization in problem (5) is denoted as $\Gamma(\mathcal{G}_{\mathcal{S}},\mathcal{G}_{\mathcal{T}},\mathfrak{U})$. In (13), we update $\mathcal{G}_{\mathcal{S}}$ and $\mathcal{G}_{\mathcal{T}}$ with $\mathfrak{U}$ fixed, i.e., we solve \{$\mathcal{G}_{\mathcal{S}}^{*},\mathcal{G}_{\mathcal{T}}^{*}$\}  = argmin$\Gamma(\mathcal{G}_{\mathcal{S}},\mathcal{G}_{\mathcal{T}},\mathfrak{U})$. Since both of them have a closed-form solution, we have $\Gamma(\mathcal{G}_{\mathcal{S}}^{*},\mathcal{G}_{\mathcal{T}}^{*},\mathfrak{U}) \le \Gamma(\mathcal{G}_{\mathcal{S}},\mathcal{G}_{\mathcal{T}},\mathfrak{U})$ for any $\mathcal{G}_{\mathcal{S}},\mathcal{G}_{\mathcal{T}},\mathfrak{U}$.
Similarly, given the closed form solution of optimal $\mathfrak{U}^{*}$, we have $\Gamma(\mathcal{G}_{\mathcal{S}}^{*},\mathcal{G}_{\mathcal{T}}^{*},\mathfrak{U}^{*}) \le \Gamma(\mathcal{G}_{\mathcal{S}}^{*},\mathcal{G}_{\mathcal{T}}^{*},\mathfrak{U})$.
Therefore, the $\Gamma(\mathcal{G}_{\mathcal{S}},\mathcal{G}_{\mathcal{T}},\mathfrak{U})$ decreases monotonically and iteratively, assuring the convergence of the proposed algorithm. As shown in Section VI, the proposed algorithm achieves convergence in less than 15 iterations.

The computational complexity mainly contains two parts:  unconstrained optimization problem in (9) and orthogonal constrained problem in (15). The number of iterations for updating $\mathcal{G}_{\mathscr{D}}$ is denoted as $N_1$, while $N_2$ is the average number of iterations for updating $\mathfrak{U}$ following each trial of updating $\mathcal{G}_{\mathscr{D}}$. For simplicity, the vectorization dimensionality of original and core tensors are denoted as  $D_o = \prod_{k=1}^{M}I_{k}$ and $D_c = \prod_{k=1}^{M}J_{k}$, respectively. Firstly, the complexity of (9) consists of SVD of $\mathbf{Z}$ in (10) and matrix inverse of $\mathbf{Q}$ in (13). The corresponding complexities are $O(N_{1}D_oD_c^{2})$ and $O[N_{1}D_c(n_{\mathcal{S}}^{2}\mathrm{log}(n_{\mathcal{S}})+n_{\mathcal{T}}^{2}\mathrm{log}(n_{\mathcal{T}}))]$, respectively. Secondly, given the SVD of $\mathbf{X}_{(k)}\mathbf{U}^{(-k)}\mathbf{G}^{\mathrm{T}} \ (k=1,...,M)$ in (15), $\mathfrak{U}$ is updated solved with complexity $O(N_{1}N_{2}(\sum_{k=1}^{M}I_k J_k^{2}))$. In total, the complexity of \emph{TA} method is $O[N_{1}D_oD_c^{2}+N_{1}D_c(n_{\mathcal{S}}^{2}\mathrm{log}(n_{\mathcal{S}})+n_{\mathcal{T}}^{2}\mathrm{log}(n_{\mathcal{T}}))+N_{1}N_{2}(\sum_{k=1}^{M}I_k J_k^{2})]$.

\section{Pure Samples based Classification Improvement}
Once the projection matrices $\{ \mathbf{U}^{(1)},\mathbf{U}^{(2)},\mathbf{U}^{(3)}\}$ are computed, source and target tensors are represented as core tensors $ \mathcal{G}_{\mathcal{S}}$ and $\mathcal{G}_{\mathcal{T}}$, respectively. The predicted map of target HSI can be easily obtained by a supervised classifier.  It is notable that only part of target tensors are well exploited for domain adaptation and superpixel segmentation contributes only to tensor construction in the whole progress. In order to further exploit all target tensors and superpixel segmentation, in this section a strategy based on pure samples extraction is introduced to improve classification performance.

\begin{figure}[!t]
\setlength{\abovecaptionskip}{-5pt}
\setlength{\belowcaptionskip}{-20pt}
\centering
\includegraphics[width=0.48\textwidth]{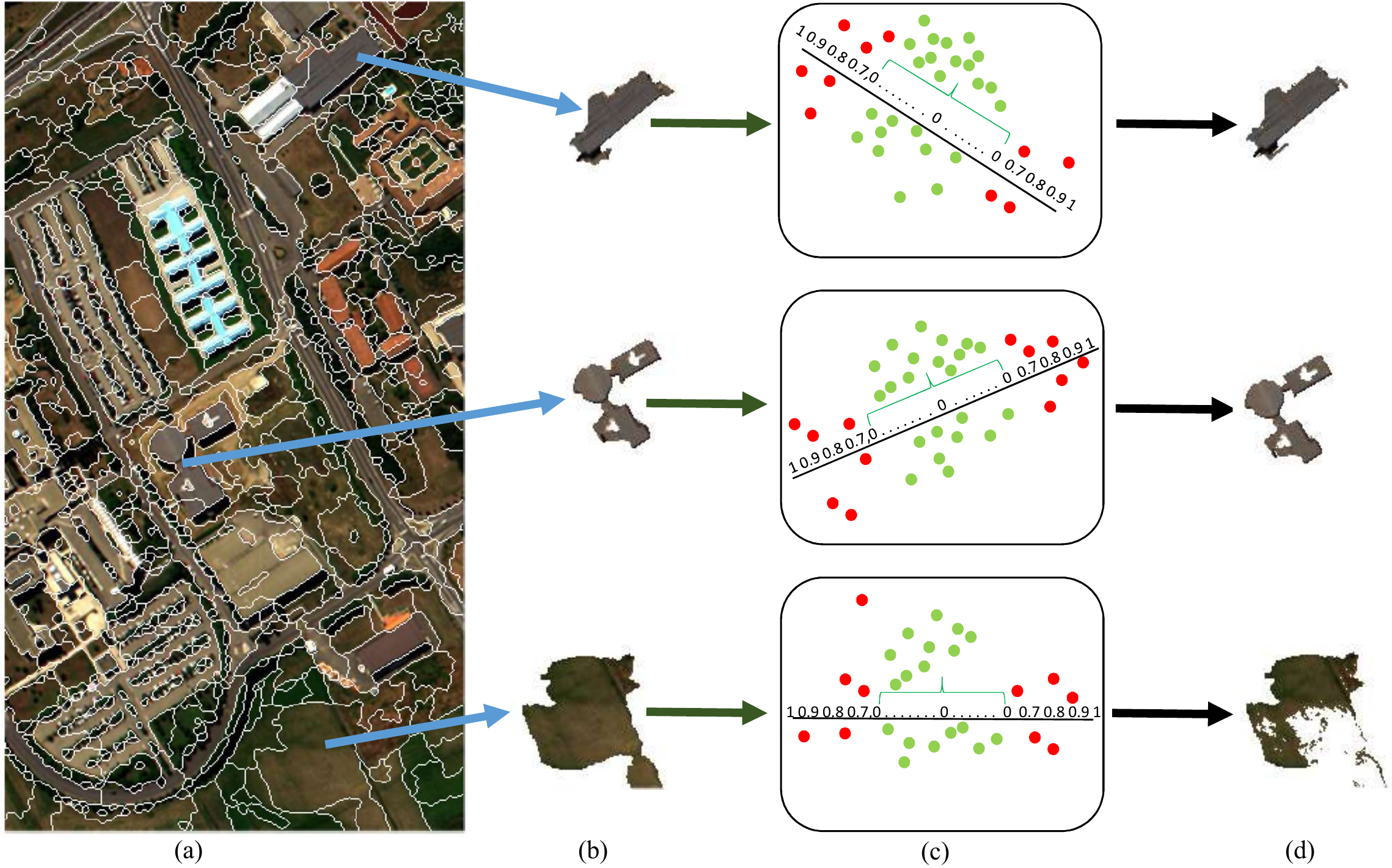}
\caption{(a) Superpixel segmentation of the Pavia University data. (b) Illustration of 3 superpixels. (c) PCA-based pure samples extraction of the 3 superpixels. Here, the min-max axis is the first principal component vector and the threshold $T$ is set as 0.7. (d) Results of pure samples for 3 superpixels. (Best viewed in color). }
\label{PCAMaxmin}
\end{figure}

We firstly introduce the PCA-based method used for extracting the pure samples.
As suggested in \cite{Martin2012}, for each superpixel we perform PCA and choose the first three principal components as the projection axis. Then, we project all samples onto these three principal components. For each projection axis, we normalize the projection values to $[0,1]$. Since samples belonging to the same class in each superpixel have similar spectral signatures, these samples are likely to be projected to middle range of $[0,1]$, instead of extreme value 0 and 1. Given a threshold $T$ (i.e., 0.9), if the normalized projection of sample $p_i$ is larger than  $T$, we assign a weight of $p_i$ to the sample. Otherwise if it is smaller than $1\!-\!T$, the weight is set as $1\!-\!p_i$. Further, 0 is assigned to those pixels which meet $p_i\in(1-T,T)$. In this way, each pixel is represented by three weights for three components.
Finally, the sum of all weights for each sample is regarded as its purity index. The samples with purity index equal to 0 are extracted as pure samples. Illustrative examples of pure samples extraction is shown in Fig. \ref{PCAMaxmin}, where $T$ is set as 0.7.
After the extraction of pure samples in target HSI, we can apply the strategies for performance improvement.

The pure samples in each superpixel are expected to belong to the same class. However, there are always some samples predicting as different class in the testing stage. Therefore, if most of pure samples in one superpixel are predicted as $i$-th class, it is reasonable to believe the residual pure samples belong to the $i$-th class. Indeed, this idea is similar to the spatial filtering which also exploits spatial consistency. Since samples in one superpixel may belong to two even more classes and there are always classification errors in DA, the ratio of pure samples predicting as the same class might be reduced if we extract
more pure samples. Therefore, we extract the pure samples by fixing the ratio of pure samples predicting as same class as $0.7$. If 70\% pure samples are predicted to belong to the $i$-th class, then remaining 30\% pure samples are changed into the $i$-th class. To find the optimal pure samples,  a greedy algorithm is applied to extract more pure samples so that the ratio is no more than $0.7$:
\begin{equation}
N_{pure}^{*}= \min\{N_{pure}|ratio(N_{pure})\leq0.7\}
\end{equation}
where $ratio(\cdot)$ means the ratio predicting as the same class.
Intuitively, the optimal $N_{pure}^{*}$ should be as large as possible (to include more samples for the purpose of improving classification). Meanwhile, it should not be too large, otherwise samples belonging to different classes are included.
We reduce the threshold $T$ by 0.01 iteratively to include more pure samples until the ratio of pure samples predicting as the same class reaches 70\%. Then predicted results of  remaining 30\% pure samples are alternated as the predicted class of the 70\% pure samples. We denote the strategy as \emph{TA\_P} for short. Although the strategy is simple, experimental results in section V reveal that remarkable margins are gained by \emph{TA\_P} over the proposed \emph{TA} method. 

\section{Data Description and Experimental Setup}
\begin{figure}[!t]
\setlength{\abovecaptionskip}{-5pt}
\setlength{\belowcaptionskip}{-20pt}
\centering
\includegraphics[width=0.5\textwidth]{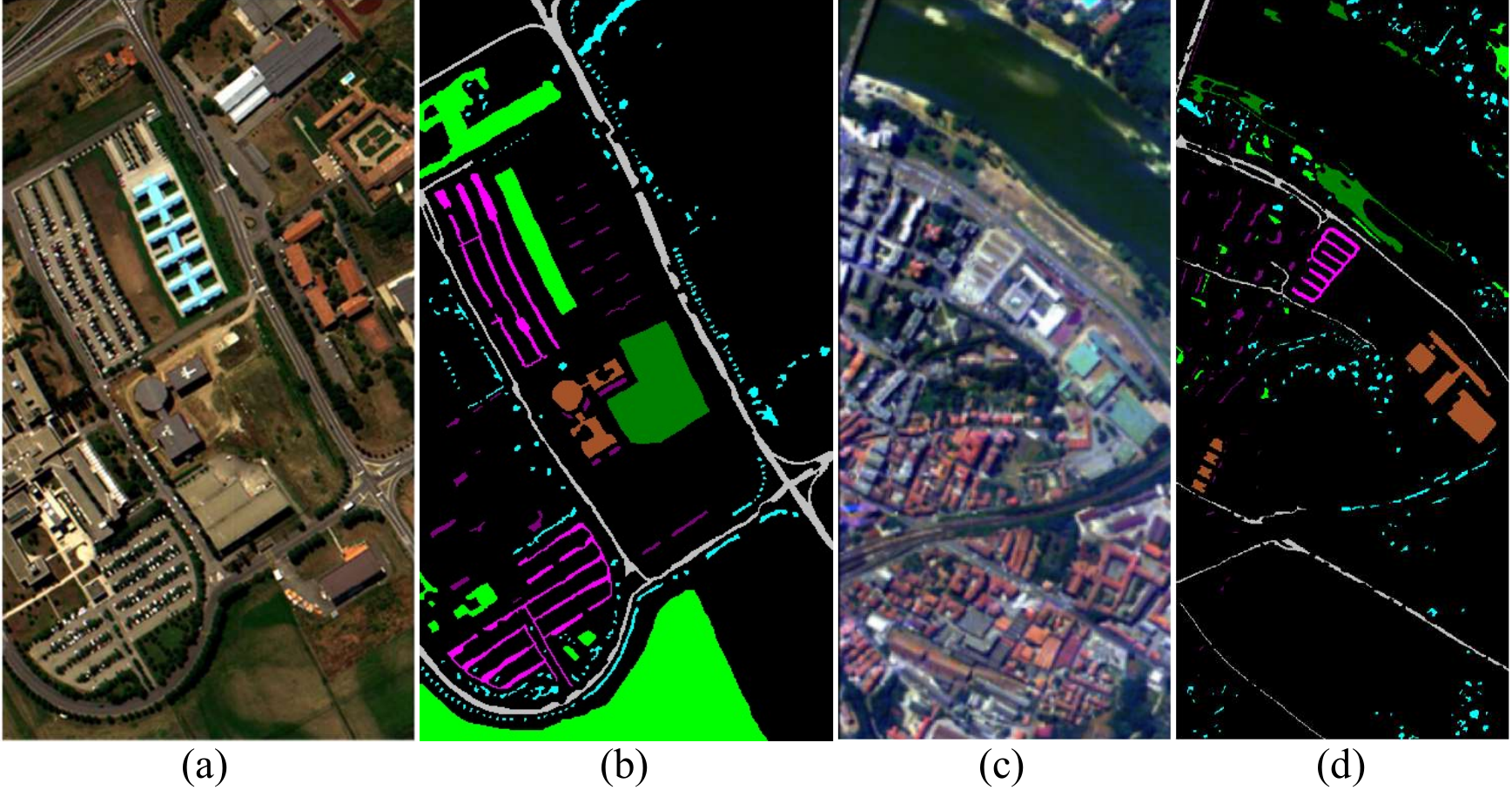}
\caption{ROSIS Pavia dataset used in our experiments. (a) Color composite image and (b) ground truth of the university scene; (c) color composite image and (d) ground truth of city center scene.}
\label{Data_Pavia}
\end{figure}

\begin{figure}[!t]
\setlength{\abovecaptionskip}{-5pt}
\setlength{\belowcaptionskip}{-20pt}
\centering
\includegraphics[width=0.5\textwidth]{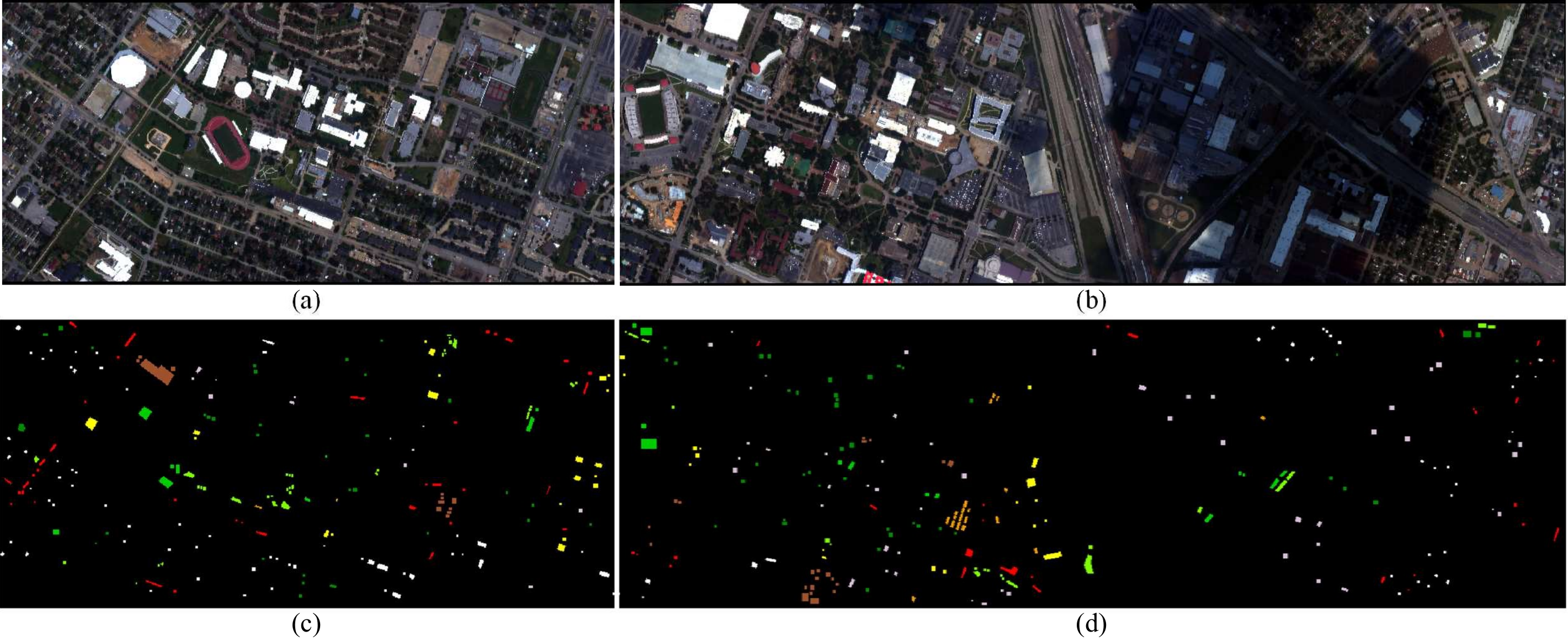}
\caption{Houston GRSS2013 dataset used in our experiments. (a) Color composite image and (b) ground truth of the left dataset; (c) color composite image and (d) ground truth of the right dataset.}
\label{Data_GRSS}
\end{figure}
\subsection{DataSet Description}
The first dataset consists in two hyperspectral images acquired by the Reflective Optics Spectrographic Image System (ROSIS) sensor over the University of Pavia and Pavia City Center are considered (see Fig. \ref{Data_Pavia}). The Pavia City Center image contains 102 spectral bands and has a size of 1096$\times$492 pixels. The Pavia University image contains instead 103 spectral reflectance bands and has a size of 610$\times$340 pixels. Seven classes shared by both images are considered in our experiments.  The number of labeled pixels available is detailed in Table \ref{Samples}. In the experiments, the Pavia University image was considered as the source domain, while the Pavia City Center image as the target domain, or vice versa. These two cases are denoted as \emph{univ}/\emph{center} and \emph{center}/\emph{univ}, respectively. Note that only 102 spectral bands of the Pavia University image were used for adaptation.

The second dataset is the GRSS2013 hyperspectral image consisting of 144 spectral bands. The data were acquired by the National Science Foundation (NSF)-funded Center for Airborne Laser Mapping (NCALM) over the University of Houston campus and the neighboring urban area. 
Originally, the data sets have a size of $1905 \times 349$ pixels and their ground truth includes 15 land cover
types. Similarly to the previous case, we consider two disjoint sub-images with $750 \times 349$ pixels
[Fig. \ref{Data_GRSS}(a)] and $1155 \times 349$ pixels [Fig. \ref{Data_GRSS}(b)], respectively. For ease of reference, we name the two cases as \emph{left}/\emph{right} and \emph{right}/\emph{left}. These sub-images share eight classes in the
ground truth: \emph{``healthy grass''}, \emph{``stressed grass''}, \emph{``trees''}, \emph{``soil''}, \emph{``residential''}, \emph{``commercial''}, \emph{``road''} and \emph{``parking lot 1''}. The classes are listed in Table \ref{Samples} with the corresponding number of samples.
  
\definecolor{class11}{RGB}{192,192,192}
\definecolor{class12}{RGB}{0,255,0}
\definecolor{class13}{RGB}{0,255,255}
\definecolor{class14}{RGB}{0,128,0}
\definecolor{class15}{RGB}{255,0,255}
\definecolor{class16}{RGB}{165,82,41}
\definecolor{class17}{RGB}{128,0,128}
\definecolor{class21}{RGB}{0, 205, 0}
\definecolor{class22}{RGB}{127, 255, 0}
\definecolor{class23}{RGB}{0, 139, 0}
\definecolor{class24}{RGB}{160, 82, 45}
\definecolor{class25}{RGB}{255, 255, 255}
\definecolor{class26}{RGB}{216, 191, 216}
\definecolor{class27}{RGB}{255, 0, 0}
\definecolor{class28}{RGB}{255, 255, 0}
\definecolor{class29}{RGB}{238, 154, 0}
\begin{table}[!t]
\setlength{\abovecaptionskip}{0pt}
\setlength{\belowcaptionskip}{0pt}
\caption{Number of Labeled Samples Available for the Pavia Dataset (top) and the GRSS2013 Dataset (down). }
\label{Samples}
{\begin{tabular}{llccc@{}}
\hline 
{No.} & { Class }&  { Color in Fig. \ref{Data_Pavia}}  & {Pavia University }& {Pavia Center} \\
\hline\hline
1     & Asphalt& \cellcolor{class11}    & 6631     &7585   \\ 
2     & Meadows& \cellcolor{class12}    & 18649        &2905 \\ 
3    & Trees& \cellcolor{class13}    & 3064           &6508 \\ 
4     & Baresoil& \cellcolor{class14}    & 5029         &6549  \\
5   & Bricks& \cellcolor{class15}    & 3682            &2140 \\
6    &Bitumen & \cellcolor{class16}    & 1330       &7287  \\
7    &Shadows & \cellcolor{class17}    & 947             &2165  \\
\hline
\hline
{No.} & { Class }&  { Color in Fig. \ref{Data_GRSS}}  & {Left  }& {Right}\\
\hline
1     & Healthy grass & \cellcolor{class21}    & 547    &704  \\
2     & Stressed grass& \cellcolor{class22}    & 569     &685 \\
3     & Trees& \cellcolor{class23}    & 451     &793\\
4     & Soil& \cellcolor{class24}    & 760     &482\\
5   & Residential& \cellcolor{class25}    & 860     &408\\
 6     & Commercial& \cellcolor{class26}    & 179     &1065 \\
7     & Road& \cellcolor{class27}    &697     &555\\
8     &Parking Lot 1& \cellcolor{class28}    & 713    &520\\
\hline
\end{tabular}}
\end{table}

\subsection{Experimental Setup}
To investigate the classification performance of the proposed methods, SVM with linear kernel is employed as the supervised classifier. In detail, it is trained on the labeled source samples and tested on the unlabeled target samples. Although classifier like SVM with Gaussian kernel performs better in the classification task, the optimal parameters of such classifier tuned by source samples usually perform worse than expected for target samples under the context of DA. On the other hand, simple linear kernel is not biased by parameter tuning and can capture original relationships between samples from different domains. Free parameter $C$ for linear SVM is tuned in the range (0.001-1000) by 5-fold cross validation.

Several unsupervised DA approaches for visual and remote sensing applications are  employed as baseline methods:\\
\begin{tabularx}{0.5\textwidth}{lX}
$\!\!\!\bullet\!\!\!$ & {\emph{Source (SRC)}: SRC is the first baseline method that trains the classifier directly utilizing the labeled source samples. } \\ 
$\!\!\!\bullet\!\!\!$& {\emph{Target (TGT)}: TGT  is the second baseline that trains the classifier directly utilizing the labeled target samples (\emph{upper} bound on performance). } \\ 
$\!\!\!\bullet\!\!\!$& {\emph{PCA}: PCA treats source and target samples as a single domain.} \\ 
$\!\!\!\bullet\!\!\!$& {\emph{GFK}: GFK proposes a closed-form solution to bridge the subspaces of the two domains using a geodesic flow on a Grassmann manifold. } \\ 
$\!\!\!\bullet\!\!\!$ & {\emph{SA}:  SA directly adopts a linear projection to match the differences between the source and target subspaces. Our approach is closely related to this method. } \\ 
$\!\!\!\bullet\!\!\!$& {\emph{TCA}: TCA carries out adaptation by learning some transfer components across domains in a RKHS using MMD.}\\
\end{tabularx}
\begin{table*}[!t]
\setlength{\abovecaptionskip}{0pt}
\setlength{\belowcaptionskip}{0pt}
\caption{Classification Results for the Pavia Dataset with Different Numbers of Labeled Source Samples. The First Three Best Results of Mean OAs For Each Column are Reported in Italic Bold, Underlined Bold and Bold, Respectively. The Proposed Approaches Outperform All the Baseline DA Methods.}
\label{result_Pavia}
\begin{tabularx}{1\textwidth}{cccXXXXXXXX}
\toprule[2pt]
\multicolumn{3}{c}{\# Labeled samples per class}  & \emph{5} & \emph{10}  & \emph{15} & \emph{20} & \emph{25} & \emph{30} & \emph{35} & \emph{40}\\
\midrule[1pt]
\multirow{16}{*}{\emph{univ}/\emph{center}} & \multirow{2}{*}{\emph{SRC}} & OA &$69.7/0.78$& $72.8/0.61$ &$73.0/0.62$ &$71.6/0.74$ & $70.0/0.71$ & $68.3/0.62$ & $67.5/0.64$ &$66.9/0.51$ \\
{} &{}                               & Kappa &$0.638/0.091$& $0.673/0.072$ &$0.675/0.073$ &$0.659/0.087$ & $0.640/0.084$ &  $0.620/0.073$ & $0.611/0.075$  &$0.605/0.060$    \\
\cline{2-11}
{} &\multirow{2}{*}{\emph{TGT}} & OA & 
$\textbf{\emph{82.1}}/0.42$& $\textbf{\emph{85.9}}/0.25$ &$\textbf{\emph{87.9}}/0.15$ &$\textbf{\emph{88.6}}/0.12$ & $\textbf{\emph{89.0}}/0.11$ & $\textbf{\emph{89.7}}/0.10$ & $\textbf{\emph{90.0}}/0.11$ &$\textbf{\emph{90.6}}/0.10$ \\
{} &{}                               & Kappa&$0.786/0.050$& $0.832/0.030$ &$0.854/0.017$ &$0.863/0.014$ & $0.868/0.013$ &  $0.877/0.012$ & $0.880/0.013$  &$0.887/0.012$ \\
\cline{2-11}
{} &\multirow{2}{*}{\emph{PCA}} & OA &$69.5/0.83$& $72.9/0.60$ &$74.6/0.49$ &$75.2/0.51$ & $76.0/0.50$ & $75.7/0.47$ & $74.7/0.51$ &$74.9/0.49$\\
{} &{}                               & Kappa&$0.634/0.098$& $0.674/0.071$ &$0.693/0.059$ &$0.701/0.061$ & $0.710/0.061$ &  $0.706/0.057$ & $0.694/0.061$  &$0.696/0.059$ \\
\cline{2-11}
{} &\multirow{2}{*}{\emph{GFK}} & OA & $66.8/0.53$& $67.0/0.47$ &$67.7/0.47$ &$68.5/0.41$ & $68.3/0.41$ & $67.7/0.34$ & $67.4/0.37$ &$67.4/0.36$\\
{} &{}                               & Kappa&$0.601/0.061$& $0.604/0.054$ &$0.612/0.055$ &$0.621/0.048$ & $0.619/0.049$ &  $0.612/0.040$ & $0.609/0.043$  &$0.610/0.041$ \\
\cline{2-11}
{} &\multirow{2}{*}{\emph{SA}} & OA & $69.0/0.81$& $72.6/0.63$ &$74.2/0.48$ &$75.0/0.50$ & $75.9/0.48$ & $75.7/0.44$ & $74.8/0.49$ &$75.2/0.48$ \\
{} &{}                               & Kappa&$0.628/0.095$& $0.670/0.074$ &$0.689/0.058$ &$0.698/0.060$ & $0.709/0.058$ &  $0.706/0.053$ & $0.695/0.059$  &$0.701/0.057$ \\
\cline{2-11}
{} &\multirow{2}{*}{\emph{TCA}} & OA &$69.4/0.73$& $73.0/0.54$ &$74.0/0.50$ &$74.8/0.49$ & $75.0/0.50$ & $74.7/0.46$ & $74.1/0.49$ &$73.6/0.49$\\
{} &{}                               & Kappa&$0.634/0.086$& $0.675/0.064$ &$0.687/0.060$ &$0.696/0.059$ & $0.698/0.060$ &  $0.694/0.056$ & $0.687/0.058$  &$0.682/0.059$  \\
\cline{2-11}
{} &\multirow{2}{*}{Proposed \bf{\emph{TA}}} & OA &$\mathbf{74.3}/0.80$& $\mathbf{79.5}/0.49$ &$\mathbf{80.7}/0.43$ &$\mathbf{81.2}/0.41$ & $\mathbf{82.1}/0.33$ & $\mathbf{81.9}/0.32$ & $\mathbf{82.4}/0.27$ &$\mathbf{83.1}/0.26$\\
{} &{}                               & Kappa&$0.691/0.094$& $0.753/0.059$ &$0.767/0.052$ &$0.774/0.050$ & $0.783/0.040$ &  $0.782/0.038$ & $0.787/0.033$  &$0.797/0.032$   \\
\cline{2-11}
{} &\multirow{2}{*}{Proposed \bf{\emph{TA\_P}}} & OA & $\underline{\mathbf{76.4}}/0.93$& $\underline{\mathbf{81.8}}/0.56$ &$\underline{\mathbf{83.2}}/0.52$ &$\underline{\mathbf{83.9}}/0.48$ & $\underline{\mathbf{84.9}}/0.40$ & $\underline{\mathbf{84.5}}/0.38$ & $\underline{\mathbf{85.0}}/0.31$ &$\underline{\mathbf{86.1}}/0.30$\\
{} &{}                               & Kappa&$0.716/0.110$& $0.780/0.068$ &$0.797/0.063$ &$0.805/0.058$ & $0.818/0.048$ &  $0.812/0.046$ & $0.818/0.037$  &$0.832/0.037$ \\
\midrule[1pt]
\multirow{16}{*}{\emph{center}/\emph{univ}} & \multirow{2}{*}{\emph{SRC}} & OA &$\mathbf{53.2}/0.72$& $\underline{\mathbf{56.9}}/0.56$ &$56.6/0.47$ &$56.8/0.44$ & $57.2/0.42$ & $57.3/0.36$ & $56.2/0.41$ &$55.9/0.42$ \\
{} &{}                               & Kappa& $0.409/0.069$& $0.448/0.053$ &$0.449/0.044$ &$0.451/0.043$ & $0.454/0.039$ &  $0.456/0.033$ & $0.446/0.038$  &$0.444/0.040$   \\
\cline{2-11}
{} &\multirow{2}{*}{\emph{TGT}} & OA &
$\textbf{\emph{60.9}}/0.98$& $\textbf{\emph{69.1}}/0.74$ &$\textbf{\emph{72.7}}/0.62$ &$\textbf{\emph{75.6}}/0.62$ & $\textbf{\emph{77.0}}/0.53$ & $\textbf{\emph{79.0}}/0.51$ & $\textbf{\emph{80.9}}/0.35$ &$\textbf{\emph{81.9}}/0.32$ \\
{} &{}                               & Kappa&$0.508/0.094$& $0.601/0.075$ &$0.644/0.068$ &$0.678/0.070$ & $0.697/0.061$ &  $0.721/0.060$ & $0.744/0.042$  &$0.758/0.038$ \\
\cline{2-11}
{} &\multirow{2}{*}{\emph{PCA}} & OA &$50.9/0.78$& $55.4/0.57$ &$54.9/0.50$ &$56.1/0.44$ & $56.5/0.35$ & $56.6/0.30$ & $56.5/0.33$ &$56.5/0.27$\\
{} &{}                               & Kappa&$0.395/0.074$& $0.440/0.056$ &$0.436/0.047$ &$0.448/0.043$ & $0.451/0.034$ &  $0.454/0.027$ & $0.453/0.033$  &$0.453/0.025$    \\
\cline{2-11}
{} &\multirow{2}{*}{\emph{GFK}} & OA &$48.7/0.77$& $51.3/0.64$ &$51.1/0.55$ &$51.8/0.50$ & $53.3/0.42$ & $52.4/0.39$ & $53.4/0.36$ &$53.7/0.34$\\
{} &{}                               & Kappa& $0.373/0.070$& $0.396/0.060$ &$0.395/0.051$ &$0.401/0.045$ & $0.417/0.040$ &  $0.406/0.036$ & $0.416/0.033$  &$0.419/0.032$\\
\cline{2-11}
{} &\multirow{2}{*}{\emph{SA}} & OA &$50.7/0.84$& $55.5/0.58$ &$55.1/0.47$ &$56.2/0.44$ & $56.3/0.32$ & $56.6/0.30$ & $56.5/0.33$ &$56.5/0.26$ \\
{} &{}                               & Kappa&$0.393/0.079$& $0.441/0.057$ &$0.437/0.044$ &$0.449/0.043$ & $0.450/0.031$ &  $0.454/0.029$ & $0.453/0.033$  &$0.452/0.025$   \\
\cline{2-11}
{} &\multirow{2}{*}{\emph{TCA}} & OA & $49.7/0.71$& $55.4/0.60$ &$55.8/0.46$ &$55.9/0.45$ & $57.2/0.45$ & $57.5/0.37$ & $57.0/0.34$ &$57.1/0.35$\\
{} &{}                               & Kappa&$0.378/0.067$& $0.438/0.057$ &$0.444/0.041$ &$0.445/0.041$ & $0.457/0.042$ &  $0.461/0.032$ & $0.457/0.030$  &$0.459/0.031$  \\
\cline{2-11}
{} &\multirow{2}{*}{Proposed \bf{\emph{TA}}} & OA &$52.8/0.80$& $55.4/0.62$ &${\mathbf{56.6}}/0.47$ &$\underline{\mathbf{57.8}}/0.43$ & $\underline{\mathbf{58.4}}/0.38$ & $\underline{\mathbf{58.2}}/0.37$ & $\underline{\mathbf{59.2}}/0.33$ &$\underline{\mathbf{59.1}}/0.36$\\
{} &{}                               & Kappa&$0.416/0.074$& $0.441/0.060$ &$0.456/0.045$ &$0.465/0.041$ & $0.473/0.038$ &  $0.471/0.037$ & $0.482/0.032$  &$0.480/0.035$   \\
\cline{2-11}
{} &\multirow{2}{*}{Proposed \bf{\emph{TA\_P}}} & OA &$\underline{\mathbf{53.4}}/0.87$& $\mathbf{55.5}/0.70$ &$\underline{\mathbf{56.7}}/0.55$ &$\mathbf{57.7}/0.51$ & $\mathbf{58.2}/0.47$ & $\mathbf{57.7}/0.48$ & $\mathbf{59.0}/0.42$ &$\mathbf{58.7}/0.45$
 \\
{} &{}                               & Kappa& $0.423/0.082$& $0.444/0.069$ &$0.459/0.054$ &$0.465/0.050$ & $0.472/0.048$ &  $0.467/0.049$ & $0.481/0.043$  &$0.476/0.046$ \\
\bottomrule[2pt]
\end{tabularx}
\end{table*}
The parameters of \emph{GFK}, \emph{SA} and \emph{TCA} are tuned as in \cite{Gong2012}, \cite{Fernando2013} and \cite{Pan2011}, respectively. The dimension of final features in \emph{PCA} is set as same as \emph{SA}. The main parameters of the \emph{TA} method are the window size, the tensor dimensionality after MPCA, the core tensor dimensionality after \emph{TA} and the manifold regularization term $\lambda$.  They are fixed as $5 \times 5$ pixels, $5 \times 5 \times 20$, $1 \times 1 \times 10$ and 1e-3 in all experiments, respectively. Note that spectral dimensionality setting as 20 and spatial dimensionality unchanged in MPCA guarantee that 99\% energy is preserved and spatial information is also well kept, respectively. 

Given the computation cost of \emph{TA}, we explore the adaptation ability of \emph{TA} and \emph{TA\_P} with limited samples by randomly selecting tensors in both domains in each trial. To be specific, different numbers of tensors ([5 10 15 20 25 30 35 40] for Pavia dataset and [3 4 5 6 7 8 9 10] for GRSS2013 dataset) per class from the source domain, and 100 tensors per class from the target domain are randomly selected for adaptation. After obtaining the projection matrices, SVM classifier with linear kernel is trained on selected source tensors and tested on \emph{all} unlabeled target tensors. 
Regarding the \emph{SRC} and \emph{TGT} methods, central samples in selected source tensors and same number of samples per class randomly selected from target domain are employed for training, respectively.
In the setting of other DA baseline methods,  source tensors are vectorized as source samples and all target samples are used for adaptation. In the training stage, only the central samples are used as labeled. Take the case of 10 per class of source tensors as a example, then 250 ($5\times5\times10$) source samples per class are available for adaptation in DA baselines, and only 10 per class of central source samples are used for training the classifier. 
For each setting with same number of labeled source samples, 100 trials of the classification have been performed to ensure stability of the results. The classification results are evaluated using Overall Accuracy (OA), Kappa statistic and F-measure (harmonic mean of user's and producer's accuracies).
All our experiments have been conducted by using Matlab R2017b in a desktop PC equipped with an Intel Core i5 CPU (at 3.1GHz) and 8GB of RAM.

\section{Results and Discussions}
\subsection{Classification Performances}
\begin{table*}[!t]
\setlength{\abovecaptionskip}{0pt}
\setlength{\belowcaptionskip}{0pt}
\caption{Classification Results for the GRSS2013 Dataset with Different Numbers of Labeled Source Samples. The First Three Best Results of Mean OAs For Each Column are Reported in Italic Bold, Underlined Bold and Bold, Respectively.}
\label{result_GRSS}
\begin{tabularx}{1\textwidth}{cccXXXXXXXX}
\toprule[2pt]
\multicolumn{3}{c}{\# Labeled samples per class}  & \emph{3} & \emph{4}  & \emph{5} & \emph{6} & \emph{7} & \emph{8} & \emph{9} & \emph{10}\\
\midrule[1pt]
\multirow{16}{*}{\emph{left}/\emph{right}} & \multirow{2}{*}{\emph{SRC}} & OA &$52.2/0.46$& $55.4/0.44$ &$56.7/0.39$ &${57.6}/0.44$ & $\mathbf{59.1}/0.30$ & $59.3/0.29$ & $\mathbf{59.4}/0.26$ &$60.3/0.30$ \\
{} &{}                               & Kappa &$0.461/0.051$& $0.497/0.049$ &$0.512/0.043$ &$0.522/0.049$ & $0.538/0.034$ &  $0.540/0.032$ & $0.542/0.029$  &$0.551/0.033$   \\
\cline{2-11}
{} &\multirow{2}{*}{\emph{TGT}} & OA &$\textbf{\emph{61.7}}/0.78$& $\textbf{\emph{68.3}}/0.45$ &$\textbf{\emph{71.2}}/0.50$ &$\textbf{\emph{73.8}}/0.39$ & $\textbf{\emph{74.9}}/0.50$ & $\textbf{\emph{76.8}}/0.33$ & $\textbf{\emph{77.2}}/0.38$ &$\textbf{\emph{78.7}}/0.37$
\\
{} &{}                               & Kappa&$0.563/0.088$& $0.638/0.051$ &$0.671/0.056$ &$0.700/0.044$ & $0.713/0.057$ &  $0.734/0.037$ & $0.739/0.043$  &$0.756/0.042$   \\
\cline{2-11}
{} &\multirow{2}{*}{\emph{PCA}} & OA &$50.6/0.28$& $52.5/0.35$ &$54.2/0.30$ &$55.2/0.32$ & $56.4/0.27$ & $57.1/0.31$ & $57.7/0.27$ &$58.5/0.30$\\
{} &{}                               & Kappa&$0.443/0.031$& $0.464/0.038$ &$0.483/0.033$ &$0.494/0.036$ & $0.508/0.030$ &  $0.515/0.034$ & $0.522/0.029$  &$0.531/0.033$  \\
\cline{2-11}
{} &\multirow{2}{*}{\emph{GFK}} & OA & $53.4/0.31$& $54.9/0.27$ &$55.6/0.23$ &$56.1/0.24$ & $56.6/0.20$ & $57.0/0.23$ & $57.4/0.19$ &$57.9/0.19$ \\
{} &{}                               & Kappa&$0.474/0.035$& $0.492/0.030$ &$0.499/0.026$ &$0.504/0.027$ & $0.511/0.023$ &  $0.515/0.026$ & $0.520/0.021$  &$0.524/0.022$   \\
\cline{2-11}
{} &\multirow{2}{*}{\emph{SA}} & OA &$50.1/0.25$& $51.8/0.32$ &$53.4/0.30$ &$54.6/0.31$ & $55.9/0.26$ & $56.4/0.30$ & $56.9/0.28$ &$57.8/0.29$ \\
{} &{}                               & Kappa&$0.437/0.027$& $0.456/0.035$ &$0.474/0.033$ &$0.487/0.034$ & $0.502/0.029$ &  $0.508/0.033$ & $0.513/0.031$  &$0.523/0.032$  \\
\cline{2-11}
{} &\multirow{2}{*}{\emph{TCA}} & OA &$53.1/0.51$& $55.3/0.60$ &$56.8/0.46$ &$\underline{\mathbf{58.2}}/0.37$ & $\underline{\mathbf{59.3}}/0.37$ & $\underline{\mathbf{60.2}}/0.34$ & $\underline{\mathbf{60.0}}/0.31$ &$\underline{\mathbf{61.1}}/0.25$\\
{} &{}                               & Kappa&$0.468/0.056$& $0.493/0.065$ &$0.509/0.051$ &$0.524/0.040$ & $0.536/0.041$ &  $0.546/0.038$ & $0.544/0.034$  &$0.556/0.028$   \\
\cline{2-11}
{} &\multirow{2}{*}{Proposed \bf{\emph{TA}}} & OA &$\mathbf{54.1}/0.47$& $\mathbf{55.9}/0.46$ &$\mathbf{57.2}/0.44$ &${57.2}/0.41$ & $58.2/0.38$ & $59.5/0.36$ & $58.8/0.37$ &$60.4/0.33$\\
{} &{}                               & Kappa&$0.485/0.054$& $0.504/0.054$ &$0.520/0.052$ &$0.521/0.049$ & $0.532/0.045$ &  $0.546/0.040$ & $0.540/0.042$  &$0.557/0.038$ \\
\cline{2-11}
{} &\multirow{2}{*}{Proposed \bf{\emph{TA\_P}}} & OA &$\underline{\mathbf{54.4}}/0.49$& $\underline{\mathbf{56.1}}/0.49$ &$\underline{\mathbf{57.5}}/0.47$ &$\mathbf{57.7}/0.44$ & $58.6/0.40$ & $\mathbf{59.9}/0.36$ & $59.3/0.38$ &$\mathbf{60.9}/0.34$ \\
{} &{}                               & Kappa& $0.485/0.053$& $0.504/0.051$ &$0.518/0.048$ &$0.519/0.046$ & $0.529/0.042$ &  $0.543/0.038$ & $0.535/0.040$  &$0.552/0.036$ \\
\midrule[1pt]
\multirow{16}{*}{\emph{right}/\emph{left}} & \multirow{2}{*}{\emph{SRC}} & OA &$73.5/0.42$& $75.4/0.46$ &$76.9/0.36$ &$78.0/0.27$ & $78.0/0.30$ & $78.7/0.24$ & $79.5/0.23$ &$79.7/0.23$ \\
{} &{}                               & Kappa&$0.692/0.048$& $0.714/0.054$ &$0.732/0.042$ &$0.744/0.031$ & $0.744/0.035$ &  $0.753/0.028$ & $0.762/0.027$  &$0.764/0.027$ \\
\cline{2-11}
{} &\multirow{2}{*}{\emph{TGT}} & OA&
$\textbf{\emph{79.4}}/0.47$& $\textbf{\emph{82.7}}/0.39$ &$\textbf{\emph{83.5}}/0.38$ &$\textbf{\emph{85.2}}/0.30$ & $\textbf{\emph{85.9}}/0.32$ & $\textbf{\emph{87.2}}/0.26$ & $\textbf{\emph{87.8}}/0.30$ &$\textbf{\emph{88.8}}/0.24$ \\
{} &{}                               & Kappa&$0.760/0.055$& $0.799/0.045$ &$0.808/0.044$ &$0.827/0.035$ & $0.836/0.038$ &  $0.851/0.030$ & $0.858/0.035$  &$0.869/0.028$\\
\cline{2-11}
{} &\multirow{2}{*}{\emph{PCA}} & OA &$72.2/0.59$& $74.3/0.50$ &$76.8/0.53$ &$77.6/0.37$ & $78.1/0.39$ & $79.0/0.38$ & $80.0/0.31$ &$80.3/0.31$\\
{} &{}                               & Kappa&$0.676/0.068$& $0.701/0.058$ &$0.730/0.061$ &$0.739/0.043$ & $0.745/0.044$ &  $0.756/0.044$ & $0.767/0.036$  &$0.770/0.036$    \\
\cline{2-11}
{} &\multirow{2}{*}{\emph{GFK}} & OA &$72.0/0.55$& $72.6/0.45$ &$74.6/0.44$ &$74.9/0.41$ & $75.8/0.36$ & $75.9/0.36$ & $76.6/0.32$ &$77.4/0.27$\\
{} &{}                               & Kappa& $0.675/0.063$& $0.682/0.052$ &$0.706/0.051$ &$0.709/0.047$ & $0.720/0.042$ &  $0.721/0.042$ & $0.729/0.037$  &$0.738/0.031$   \\
\cline{2-11}
{} &\multirow{2}{*}{\emph{SA}} & OA &$71.5/0.61$& $73.5/0.51$ &$76.3/0.51$ &$76.8/0.41$ & $78.0/0.40$ & $78.7/0.39$ & $79.5/0.35$ &$80.0/0.30$ \\
{} &{}                               & Kappa&$0.669/0.070$& $0.692/0.059$ &$0.725/0.059$ &$0.730/0.047$ & $0.744/0.046$ &  $0.753/0.045$ & $0.762/0.041$  &$0.767/0.035$   \\
\cline{2-11}
{} &\multirow{2}{*}{\emph{TCA}} & OA &$64.8/0.66$& $67.6/0.49$ &$70.1/0.57$ &$71.9/0.42$ & $73.1/0.39$ & $73.9/0.35$ & $74.7/0.34$ &$75.1/0.30$\\
{} &{}                               & Kappa&$0.591/0.075$& $0.624/0.057$ &$0.653/0.065$ &$0.674/0.048$ & $0.687/0.045$ &  $0.697/0.040$ & $0.706/0.039$  &$0.711/0.035$   \\
\cline{2-11}
{} &\multirow{2}{*}{Proposed \bf{\emph{TA}}} & OA &$\mathbf{74.2}/0.65$& ${\mathbf{77.3}}/0.47$ &${\mathbf{78.8}}/0.53$ &${\mathbf{80.4}}/0.37$ & ${\mathbf{80.7}}/0.41$ & ${\mathbf{81.5}}/0.35$ & ${\mathbf{82.6}}/0.31$ &${\mathbf{83.0}}/0.31$\\
{} &{}                               & Kappa&$0.701/0.075$& $0.736/0.054$ &$0.753/0.061$ &$0.772/0.043$ & $0.776/0.047$ &  $0.785/0.040$ & $0.798/0.036$  &$0.802/0.036$    \\
\cline{2-11}
{} &\multirow{2}{*}{Proposed \bf{\emph{TA\_P}}} & OA &$\underline{\mathbf{74.6}}/0.66$& $\underline{\mathbf{77.4}}/0.49$ &$\underline{\mathbf{78.9}}/0.54$ &$\underline{\mathbf{80.5}}/0.39$ & $\underline{\mathbf{80.8}}/0.42$ & $\underline{\mathbf{81.6}}/0.36$ & $\underline{\mathbf{82.7}}/0.33$ &$\underline{\mathbf{83.2}}/0.33$ \\
{} &{}                               & Kappa& $0.705/0.076$& $0.737/0.057$ &$0.754/0.062$ &$0.774/0.044$ & $0.777/0.049$ &  $0.786/0.041$ & $0.799/0.037$  &$0.804/0.037$  \\
\bottomrule[2pt]
\end{tabularx}
\end{table*}
Tables \ref{result_Pavia} and \ref{result_GRSS} illustrate the OAs, Kappa statistics and the corresponding standard errors obtained by the proposed methods and the baseline methods for the Pavia and GRSS2013 datasets, respectively. In total, there are four cases: \emph{univ}/\emph{center}, \emph{center}/\emph{univ}, \emph{left}/\emph{right} and \emph{right}/\emph{left}.  
\subsubsection{Pavia Dataset}
Since more samples are used for adaptation and classifier training when increasing the number of labeled source samples from 5 to 40, the mean OAs and Kappa statistics  of all methods roughly increase as expected. The increasing trend of mean OAs confirms that 100 trials are enough for achieving stable results.
 Moreover, standard errors of both OAs and Kappa statistics for small numbers of labeled samples appear to be higher.
The mean OAs of \emph{TA} and \emph{TA\_P} in \emph{univ}/\emph{center} case with various number of labeled samples are in the range of 74.3\%-83.1\% and 76.4\%-86.1\%, respectively. 
However, the performance in \emph{center}/\emph{univ} case become worse for the proposed methods, with mean OA in the range of 52.8\%-59.2\% and 53.4\%-58.7\%, respectively.
Similar trend can be found for other baseline methods. These results are not a surprise: the knowledge in Pavia University data can be easily transferred to Pavia Center data, whereas it's not the same reversely.
The mean OAs achieved by \emph{TGT} for two cases are in the range of 82.1\%-90.6\% and 60.9\%-81.9\%, respectively. 
Roughly, the performance of \emph{TA\_P} in both cases is better than other baseline methods except \emph{TGT}. More specifically, \emph{TGT} yields 4.5\%-5.7\% and 7.5\%-22.0\% higher mean OAs than the \emph{TA\_P} for two cases, depending on the number of labeled samples.  
Compared with \emph{TA}, mean OA achieved by \emph{TA\_P} are on average $\sim$2.6\% higher and $\sim$0.1\% lower for the \emph{univ}/\emph{center} and \emph{center}/\emph{univ} case, respectively. 
Although the improvement for \emph{univ}/\emph{center} case is not so remarkable, this result confirms that introducing spatial consistency improves classification accuracy when the accuracy obtained by \emph{TA} is not so small.

The accuracies achieved by \emph{SRC} are about 4.6\%-16.2\% and 6.7\%-19.2\% lower than \emph{TA} and \emph{TA\_P} methods for the \emph{univ}/\emph{center} case.  
However, when numbers of labeled samples per class is no less than 15, the differences between \emph{SRC} and the proposed methods become lower for the \emph{center}/\emph{univ} case, i.e. 0\%-3.2\% and 0.1\%-3.2\% for \emph{TA} and \emph{TA\_P}, respectively.  
Further, \emph{SRC} even outperforms \emph{TA} and \emph{TA\_P} methods  with numbers of labeled samples smaller than 15.
One can see that the improvement of the proposed methods are relevant with number of labeled samples. The reason is that the adaptation ability of \emph{TA} can be enhanced as expected with more samples.
We can further notice that three methods (\emph{PCA}, \emph{GFK} and \emph{SA}) perform better with more labeled samples for both cases. When comparing them with \emph{SRC}, they roughly outperform \emph{SRC} for two cases with enough labeled samples, whereas they perform worse with a small amount of labeled samples. These observations suggest that adaptation abilities of \emph{PCA}, \emph{GFK} and \emph{SA} are severely hindered with a small number of source samples. The two proposed methods always deliver higher classification accuracies than all these three methods.  As compared with \emph{TCA}, the proposed \emph{TA\_P} method achieve higher mean OAs (i.e. 5.8\%-11.3\% and 2.0\%-3.0\% for the two cases). Since \emph{TCA} seeks a new space where domain distances are globally minimized, poor performances are achieved by \emph{TCA} with a small number of source samples, whereas the accuracies of \emph{TCA} are better than other DA methods when increasing the number of labeled samples (see last column in Table \ref{result_Pavia} for the \emph{center}/\emph{univ} case). It can be concluded that DA methods based on global alignment are affected by the availability of source samples. 

\begin{figure*}[!t]
\setlength{\abovecaptionskip}{-5pt}
\setlength{\belowcaptionskip}{-20pt}
\centering
\includegraphics[width=1\textwidth]{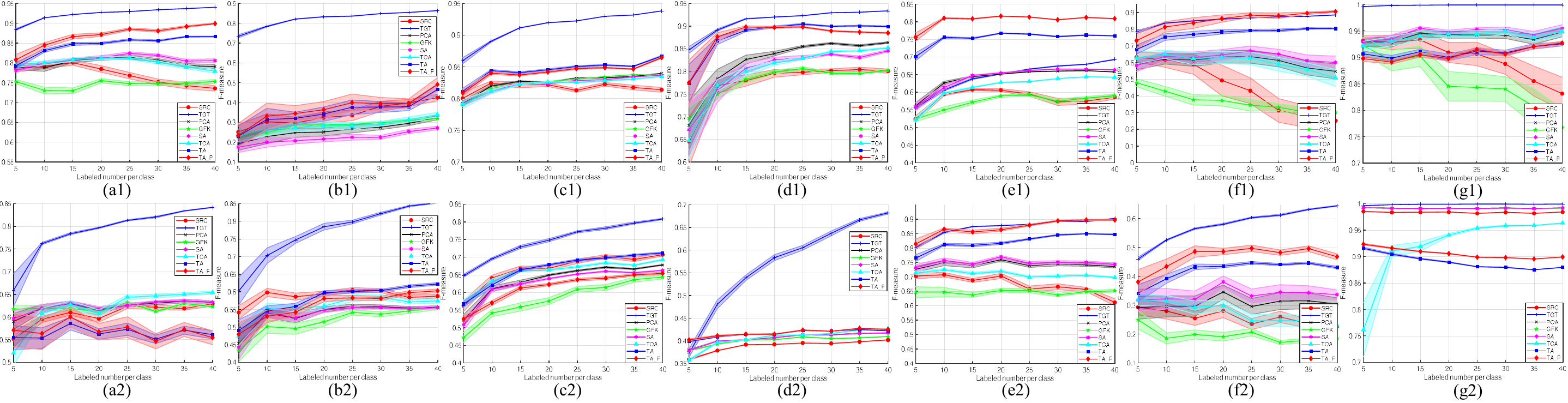}
\caption{Individual class accuracies (mean F-measure and standard errors) of different class (a-g) on the Pavia dataset. Top-row corresponds to the  \emph{univ}/\emph{center} case, while  down-row corresponds to the \emph{center}/\emph{univ} case (Best viewed in color).}
\label{F_Pavia}
\end{figure*}
\begin{figure*}[!t]
\setlength{\abovecaptionskip}{-5pt}
\setlength{\belowcaptionskip}{-20pt}
\centering
\includegraphics[width=1\textwidth]{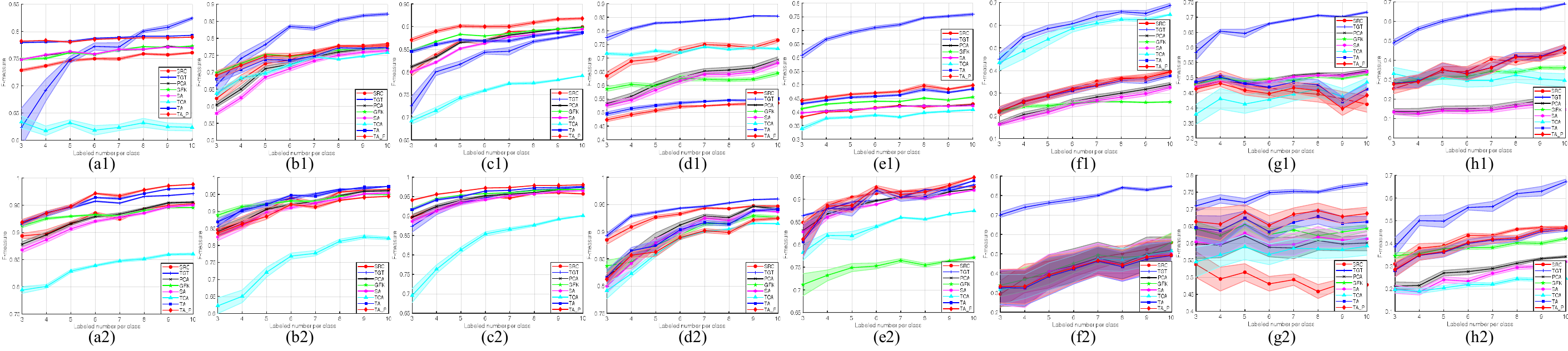}
\caption{Individual class accuracies (mean F-measure and standard errors) of different class (a-h) on the GRSS2013 dataset. Top-row represents the \emph{left}/\emph{right} case, while  down-row represents to the \emph{right}/\emph{left} case (Best viewed in color).}
\label{F_GRSS}
\end{figure*}
\begin{figure}[!t]
\setlength{\abovecaptionskip}{-5pt}
\setlength{\belowcaptionskip}{-20pt}
\centering
\includegraphics[width=0.5\textwidth]{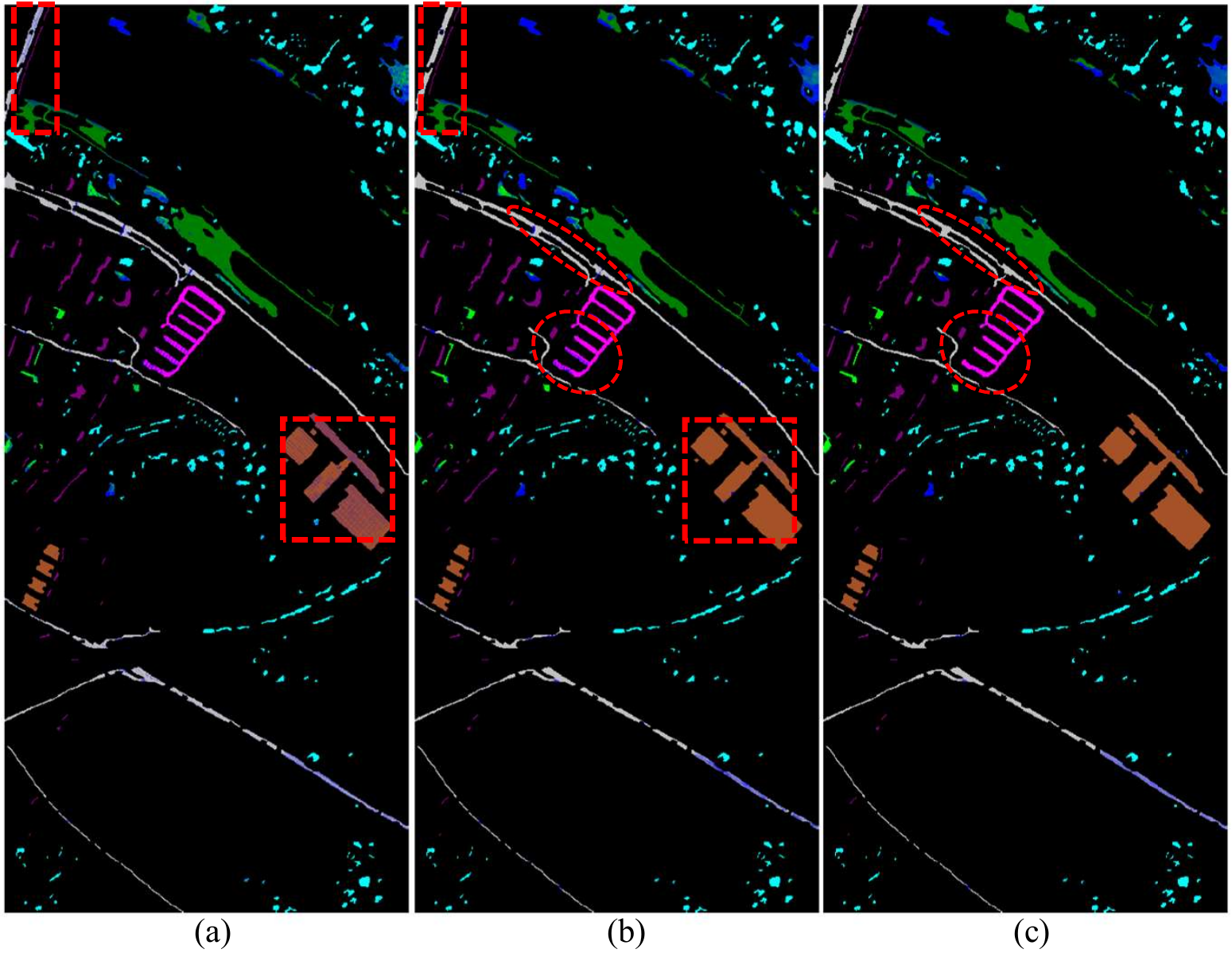}
\caption{llustration of mean classification results for \emph{univ}/\emph{center} case by (a) \emph{SRC} (Kappa = 0.605), (b) \emph{TA} (Kappa = 0.797) and (c) \emph{TA\_P} (Kappa = 0.832).}
\label{Map_Univ}
\end{figure}
\subsubsection{GRSS2013 Dataset}
Similar to the results of the Pavia dataset, the mean OAs and Kappa statistics of all methods increase as expected by increasing the number of labeled source samples, which further proves experimental stability.
The mean OAs of \emph{TA} and \emph{TA\_P} methods for \emph{left}/\emph{right} case with various numbers of labeled samples are in the range of 54.1\%-60.4\% and 54.4\%-60.9\%, respectively. 
However, both methods perform better in \emph{left}/\emph{right} case, with mean OAs in the range of of 74.2\%-83.0\% and 74.6\%-83.2\%.
By comparing classification performances of all methods,  it is clear that  GRSS2013 right data is easily transferred to left data. In fact, the difficulty of DA in \emph{left}/\emph{right} case lies in the shadow samples of the right dataset [see shadows in Fig. \ref{Data_GRSS}(b)]. 
The \emph{TA\_P} method in both cases outperforms most baseline DA methods, while \emph{TGT} achieves the best accuracy for both cases
Compared with \emph{TA}, mean OAs achieved by \emph{TA\_P} are averagely $\sim$0.4\% and $\sim$0.1\% higher for the \emph{left}/\emph{right} and \emph{right}/\emph{left} cases, respectively. 

The accuracies achieved by \emph{SRC} are $\sim$0.2\% and $\sim$0.4\% smaller than those of \emph{TA} and \emph{TA\_P} methods for the \emph{left}/\emph{right} case, respectively.  
Further, the differences become higher for the \emph{right}/\emph{left} case, i.e. 0.7\%-3.3\% and 1.1\%-3.5\% for \emph{TA} and \emph{TA\_P}, respectively.  
Both \emph{TA} and \emph{TA\_P} methods deliver higher classification accuracies than \emph{PCA}, \emph{GFK} and \emph{SA} methods. It is further observed that \emph{TCA} performs differently for the two cases, i.e. better than \emph{SRC} for \emph{left}/\emph{right} case, while worse for \emph{right}/\emph{left} case. To be specific, mean OAs achieved by \emph{TCA} are 0.5\%-0.9\% higher and 4.8\%-8.7\% lower than those obtained by \emph{SRC} for the two cases, respectively. Since MMD is a statistical distance measure of different domains,  it is reasonable that \emph{TCA} performs unstably with different cases under the setting of limited source samples. The unstable performances of \emph{TCA} are also analyzed in the next section in terms of individual class accuracies. 

\subsection{Classification Maps and Individual Class Accuracies}
To illustrate the effectiveness of the proposed \emph{TA} and \emph{TA\_P} methods, Fig. \ref{Map_Univ} provides a comparison of mean classification maps over 100 trials obtained by the \emph{SRC} and the proposed methods referring to the \emph{univ}/\emph{center} case with 40 labeled samples per class. In order to better display the classification results, average accuracy of each sample in the target image is computed and then transformed to the transparency. For example, when the average accuracy over 100 trials is 1, its  transparency is set as 0. On the other hand,  the transparency is set as 1 if the sample is never correctly classified. When blue color is applied as background in Fig. \ref{Map_Univ}, deeper of blueness for one sample means being classified less accurately.  
Compared with \emph{SRC} [Fig. \ref{Map_Univ}(a)], local classification improvement of \emph{TA} is easily observed [Fig. \ref{Map_Univ}(b)]. The ``\emph{asphalt}'' samples in the top-left (see the red rectangular box) and ``\emph{bitumen}''  samples in the right-middle (see the square box) of Pavia center data are better classified by \emph{TA} than \emph{SRC}.
Similarly, when comparing \emph{TA} with \emph{TA\_P} [Fig. \ref{Map_Univ}(c)], one can see that classification improvement of  ``\emph{asphalt}'' and ``\emph{shadows}''  classes are achieved by \emph{TA\_P}  [see red ellipses in Fig. \ref{Map_Univ}(b-c)]. 
\begin{figure*}[!t]
\setlength{\abovecaptionskip}{-5pt}
\setlength{\belowcaptionskip}{-0pt}
\centering
\includegraphics[width=1\textwidth]{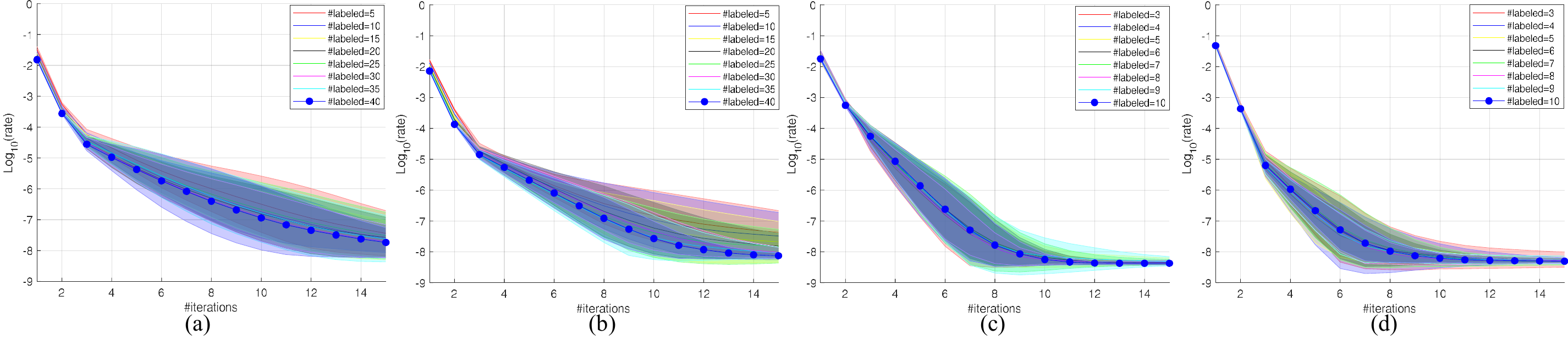}
\caption{Convergence curves of the proposed \emph{TA} method in four cases with different number of labeled samples: (a) \emph{univ}/\emph{center}, (b) \emph{center}/\emph{univ}, (c) \emph{left}/\emph{right}, (d) \emph{right}/\emph{left} (Best viewed in color).}
\label{Convergence}
\end{figure*}
\begin{figure*}[!t]
\setlength{\abovecaptionskip}{-5pt}
\setlength{\belowcaptionskip}{-20pt}
\centering
\includegraphics[width=1\textwidth]{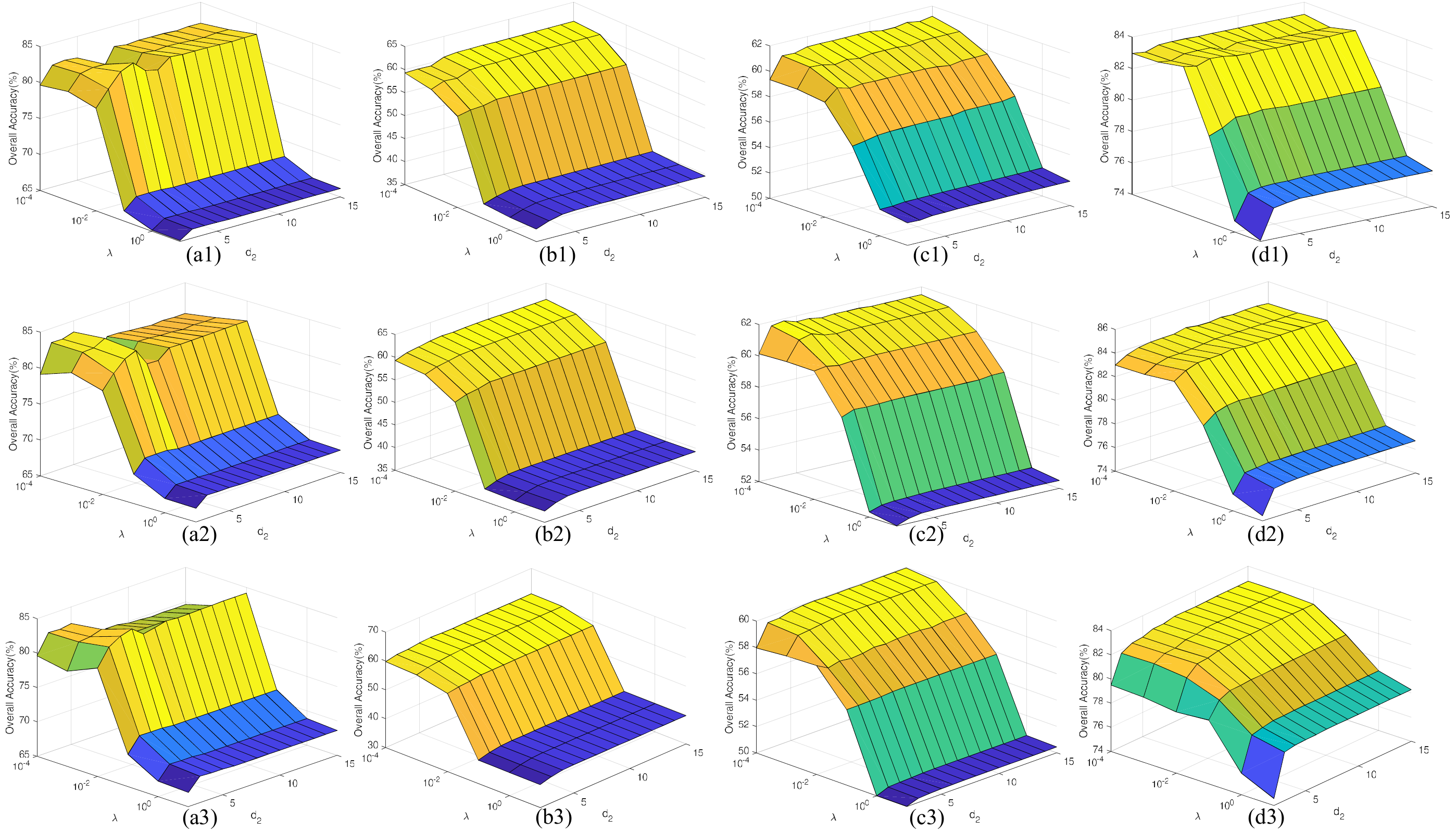}
\caption{Overall performance accuracy with different $d_2$ and $\lambda$ settings for four test cases: (a1-a3) \emph{univ}/\emph{center}; (b1-b3) \emph{center}/\emph{univ}; (c1-c3) \emph{left}/\emph{right}; (d1-d3) \emph{right}/\emph{left}. Each column corresponds to results on the same case with different window size (up to down: 3$\times$3, $5 \times 5$, 7$\times$7 pixels) (Best viewed in color).}
\label{Para}
\end{figure*}

Fig. \ref{F_Pavia} reports  individual class accuracies for the Pavia dataset, assessed by the mean F-measure (main curve) and its standard error (shaded area for each curve) over 100 trials. The results of 7 classes (``\emph{asphalt}'', ``\emph{meadows}'', ``\emph{trees}'', ``\emph{baresoil}'', ``\emph{bricks}'', ``\emph{bitumen}'', ``\emph{shadows}'') are shown in Fig. \ref{F_Pavia}(a-g), respectively. The top-row (a1-g1) corresponds to the \emph{univ}/\emph{center} case, while down-row (a2-g2) corresponds to the \emph{center}/\emph{univ} case. The following observations can be easily done:\\
$\bullet$ All methods perform more stably (smaller standard errors) with more labeled source samples in both cases. Roughly, most classes are classified more accurately with more labeled source samples except ``\emph{bitumen}' and ``\emph{shadows}' in \emph{univ}/\emph{center} case [see g(1) and f(1)]. The  ``\emph{bitumen}' in both cases and  ``\emph{trees}' in \emph{center}/\emph{univ} case have lowest accuracies than other classes, while ``\emph{meadows}'' class in both cases has the best accuracy. \\
$\bullet$ The \emph{TGT} method yields best results than other approaches for ``\emph{asphalt}'', ``\emph{meadows}'', ``\emph{trees}'', ``\emph{bitumen}'' and``\emph{shadows}'' classes, whereas fails in detecting ``\emph{bricks}'' class for \emph{univ}/\emph{center} case [see e(1)].\\
$\bullet$ The \emph{GFK} method performs the worst for most classes in both cases, and has largest standard errors for ``\emph{shadows}'' class in \emph{univ}/\emph{center} case. The  \emph{PCA} and \emph{SA} methods behave similarly over all classes, whereas \emph{SRC} performs worse for ``\emph{bitumen}'' and ``\emph{shadows}'' classes in \emph{univ}/\emph{center} case with more labeled samples.\\
$\bullet$ The \emph{TA} and \emph{TA\_P} methods outperform other DA baselines on ``\emph{trees}', ``\emph{baresoil}'', ``\emph{bricks}'' and ``\emph{bitumen}'' classes, and show comparable results on other classes, yielding a better overall classification accuracy. In addition, one can easily notice that a marginal yet obvious improvement by \emph{TA\_P} than \emph{TA} on several classes, such as ``\emph{bricks}'' and ``\emph{bitumen}''.

Referring to the GRSS2013 dataset, Fig. \ref{F_GRSS} illustrates  individual class accuracies, where top-row [a(1)-h(1)] and down-row [a(2)-h(2)]  represent \emph{left}/\emph{right} and \emph{right}/\emph{left} case, respectively. The results of 8 classes (``\emph{healthy grass}'', ``\emph{stressed grass}'', ``\emph{tress}'', ``\emph{soil}'', ``\emph{residential}'', ``\emph{commerical}'', ``\emph{road}'', ``\emph{parking lot1}'') are shown in Fig. \ref{F_GRSS}(a-h), respectively. Similarly, conclusions related to the two proposed methods can be drawn. The accuracies of ``\emph{healthy grass}'',  ``\emph{tress}'' and ``\emph{residential}'' achieved by \emph{TA} and \emph{TA\_P} are higher than other DA baselines. \emph{TCA} outperforms most methods on ``\emph{commerical}'' class for \emph{left}/\emph{right} case. Considering that ``\emph{commerical}'' is in the shadow area of right dataset, it is concluded that MMD measure is more suitable for large domain divergenve. 

\subsection{Convergence}
We have proved that the \emph{TA} method is convergent under the iteratively updating rules of
projection matrices and core tensors. Here, we investigate and demonstrate the speed of the convergence
based on experimental results.
Fig. \ref{Convergence} shows the mean reducing rate of the objective function of \emph{TA} on the four test cases over 100 trials (shaded areas represent standard errors).
One can see that the objective function value decreases by increasing the number of iterations. Moreover, we can observe that the \emph{TA} converges very fast, usually taking less than 15 iterations. The reducing rate reaches 10e-6 and 10e-7 for the Pavia and the GRSS2013 datasets over all the 100 trials at the 15-\emph{th} iteration, respectively. For our MATLAB implementation, 15 iterations take $\sim$3.3s for \emph{univ}/\emph{center} cases when 40  labeled samples per class are used.
\subsection{Parameter Sensitivity Analysis}
Herein we perform experiments to discuss effects of parameters of the \emph{TA} methods in the four cases.
For simple and valid quantitative analysis, we only employ mean  OAs to evaluate different parameter configurations. In detail, window size, spectral dimensionality of \emph{TA} $d_2$ and manifold regularization $\lambda$ are discussed to achieve better understanding of the proposed method. Assuming that the window size is $W \times W$ pixels, we set tensor dimensions for MPCA and \emph{TA} as $W \times W \times 20$ and $1 \times 1 \times d_2$, respectively. Note that the number of labeled samples are fixed to 40 per class  for Pavia dataset and 10 per class for GRSS2013 dataset, and 10 trials are conducted for each sub-experiment. 
Fig. \ref{Para} illustrates the mean OAs with respect to different parameter configurations for the four cases. Each column indicates results using different window sizes for each case. One can observe that the trends of the mean OAs for all test cases under different window sizes are nearly the same, i.e. large value of $\lambda$ and low value of $d_2$ can both yield worse accuracy.  
The observation points out two conclusions: 1) large values of $\lambda$ force strong geometry preservation, hindering the learning of projection matrices; 2) If $d_2$ is smaller than 5 for all cases, no enough spectral information is preserved for training the classifier. However, when $d_2$ is lager than 10, there is no improvement of classification results.  To sum up, $\lambda \in [0, 0.001]$ and $d_2 \in [10, 20]$ can be optimal parameter values for all cases.

\section{Conclusion}
This paper has addressed the issue of DA in the classification of HSI under the assumption of small number of labeled samples. The main contributions of this paper are the proposed tensor alignment based domain adaptation algorithm and the strategy based on pure samples extraction for performance improvement. 

The proposed \emph{TA} naturally treats each sample in HSI as a 3-D tensor, exploiting the multilinear relationship between spatial and spectral dimensions. 
The shift between 3-D tensors from different domains is reduced by introducing a common set of projection matrices in the Tucker decomposition. 
The \emph{TA} method mainly contains three steps, i.e. tensors construction, dimension reduction and tensor alignment. Firstly, HSIs in both domains are segmented into superpixels and each tensor is constructed by including samples from the same superpixel. In this way, the tensors are expected to contain samples belonging to the same class.
Then, in order to reduce computational cost, MPCA is employed for spectral dimension reduction. In the stage of tensor alignment, to preserve the geometry of original tensors, two laplacian matrices from both domains are first computed. The problem of tensor alignment is formulated as jointly Tucker decomposition of tensors from both domains with manifold regularization on core tensors and orthogonal regularization on projection matrices.
The solution is found by the developed efficient iterative algorithm and the convergence is analyzed. 
Once the projection matrices are computed, source and target tensors are represented as core tensors. The predicted map of target HSI can be easily obtained by a supervised classifier.

To further exploit the spatial consistency of HSI, a strategy for pure samples extraction for performance improvement is then proposed. The pure samples in each superpixel have similar spectral features and likely belong to the same class. Given that samples in one superpixel may belong to two or even more classes, pure samples may include samples belonging to different classes if we increase the number of pure samples extraction. To extract an appropriate number of pure samples, we fix the ratio of pure samples (predicted as the same class) as a constant value. We consider that it is reasonable to assume 70\% pure samples in one superpixel. Although the strategy is simple, it turns out to be effective in performance improvement. 

The experiments are conducted on four real HSIs, i.e. Pavia University and City Center, GRSS left and right images. To explore the adaptation capacity of \emph{TA}, different numbers of source samples are randomly selected as labeled. Given the computational cost of \emph{TA}, 100 tensors per class from target domain are selected for alignment. The \emph{TA} method yields better results on \emph{univ/center}, \emph{center/univ} and \emph{right/left} data sets than the other considered DA methods, whereas \emph{TCA} outperforms \emph{TA} on \emph{left/right} data set. It is found that \emph{TCA} performs better than all subspace learning method on the ``\emph{Commercial}'' class in the \emph{left/right} data set. Since the MMD-based \emph{TCA} method can directly reduce the domain divergence, the ``\emph{Commercial}'' class obtained under different conditions in the \emph{left/right} data set (non-shadow and shadow area in the two images) is better adapted by \emph{TCA} than by other considered DA methods. To summarize, the proposed \emph{TA} method can achieve better performance compared with the state-of-the-art subspace learning methods when a limited amount of source labeled samples are available.

As future development, the proposed \emph{TA} method can be easily extended to manifold regularization orthogonal 
Tucker decomposition for tensor data dimension reduction. Its typical applications include multichannel electroencephalographies, multiview images and videos processing.

\section*{Acknowledgment}
The authors would like to thank Prof. P. Gamba from the
University of Pavia for providing the ROSIS data, thank the Hyperspectral Image Analysis Group and the NSF
Funded Center for Airborne Laser Mapping (NCALM) at the University
of Houston for providing the grss\_dfc\_2013 data set, and
the IEEE GRSS Data Fusion Technical Committee for organizing the
2013 Data Fusion Contest.

\appendices
\section{Proof of Solving Eq. (15)}
$\mathbf{Theorem \ 1}$  Let $ \mathbf{\Lambda DV}^{\mathrm{T}}$  be the singular value decomposition of $\mathbf{AB}^{\mathrm{T}}$, where $\mathbf{A} \in \mathbb{R}^{m \times n}$ and $\mathbf{B} \in \mathbb{R}^{p \times n}$. Then $\mathbf{X}=\mathbf{\Lambda I_{m \times p}V}^{\mathrm{T}}$  is an orthogonal matric minimizing $||\mathbf{A}-\mathbf{XB}||_{\mathrm{F}}^{2}$, where $\mathbf{I}_{m \times p}$ is a matrix with  diagonal elements all are 1 while others are 0. \\
\emph{Proof}. To derive the method we first expand $||\mathbf{A}-\mathbf{XB}||_{\mathrm{F}}^{2}$:
\begin{eqnarray}
||\mathbf{A}-\mathbf{XB}||_{\mathrm{F}}^{2}= ||\mathbf{A}||_{\mathrm{F}}^{2}+||\mathbf{B}||_{\mathrm{F}}^{2}-2\mathrm{tr}({\mathbf{X}^{\mathrm{T}}\mathbf{AB}^{\mathrm{T}}})
\end{eqnarray}
So picking $\mathbf{X}$ to maximize $\mathrm{tr}({\mathbf{X}^{\mathrm{T}}\mathbf{AB}^{\mathrm{T}}})$ will minimize $||\mathbf{A}-\mathbf{XB}||_{\mathrm{F}}^{2}$.
Let $ \mathbf{\Lambda DV}^{\mathrm{T}}$  be the singular value decomposition of $\mathbf{AB}^{\mathrm{T}}$. Then we have:
\begin{eqnarray}
\mathrm{tr}({\mathbf{X}^{\mathrm{T}}\mathbf{AB}^{\mathrm{T}}})=\mathrm{tr}(\mathbf{X}^{\mathrm{T}} \mathbf{\Lambda DV}^{\mathrm{T}})=\mathrm{tr}(\mathbf{V}^{\mathrm{T}}\mathbf{X}^{\mathrm{T}} \mathbf{\Lambda D})
\end{eqnarray}
Write $\mathbf{Z}=\mathbf{V}^{\mathrm{T}}\mathbf{X}^{\mathrm{T}} \mathbf{\Lambda}$, notice $\mathbf{Z}$ is orthogonal (being the product of orthogonal matrices).
 The goal is re-stated: maximize $\mathrm{tr}(\mathbf{ZD})$ through our choice of $\mathbf{X}$. Since $\mathbf{D}$ is  diagonal, then $\mathrm{tr}(\mathbf{ZD})=\sum_{i}\mathbf{Z}_{ii}\mathbf{D}_{ii}$. The $\mathbf{D}_{ii}$ are non-negative and $\mathbf{Z}$ is orthogonal for any choice of $\mathbf{X}$. The maximum is achieved by choosing $\mathbf{X}$ such that all of $\mathbf{Z}_{ii}=1$ which implies $\mathbf{Z}=\mathbf{I}_{m \times p}$. So an optimal $\mathbf{X}$ is $\mathbf{\Lambda I_{m \times p}V}^{\mathrm{T}}$.


\ifCLASSOPTIONcaptionsoff
  \newpage
\fi

\bibliographystyle{IEEEtran}
\bibliography{tgrs}

\begin{IEEEbiography}[{\includegraphics[width=1in,height=1.25in,clip,keepaspectratio]{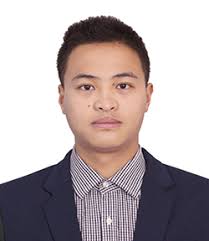}}]{Yao Qin}
(SM' 16) received the B.S degree in Information Engineering from Shanghai Jiaotong University, Shanghai, China, in 2013 and the M. S. degree from School of Electronic and Engineering, National University of Defence Technology, Changsha, China, in 2015. Currently, he is pursuing the Ph. D. degree and is a visiting Ph. D at RSLab in the Department of Information Engineering and Computer Science, University of Trento, Italy. His major research interests are remote sensing image classification and domain adaptation.
\end{IEEEbiography}
\begin{IEEEbiography}[{\includegraphics[width=1in,height=1.25in,clip,keepaspectratio]{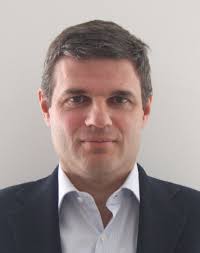}}]{Lorenzo Bruzzone}
(S'95-M'98-SM'03-F'10) received the Laurea (M.S.) degree in electronic engineering (\emph{summa cum laude}) and the Ph.D. degree in telecommunications from the University of Genoa, Italy, in 1993 and 1998, respectively. 

He is currently a Full Professor of telecommunications at the University of Trento, Italy, where he teaches remote sensing, radar, and digital communications. Dr. Bruzzone is the founder and the director of the Remote Sensing Laboratory in the Department of Information Engineering and Computer Science, University of Trento. His current research interests are in the areas of remote sensing, radar and SAR, signal processing, machine learning and pattern recognition. He promotes and supervises research on these topics within the frameworks of many national and international projects. He is the Principal
Investigator of many research projects. Among the others, he is the Principal Investigator of the \emph{Radar for icy Moon exploration} (RIME) instrument in the framework of the \emph{JUpiter ICy moons Explorer} (JUICE) mission of the European Space Agency. He is the author (or coauthor) of 215 scientific publications in referred international journals (154 in IEEE journals), more than 290 papers in conference proceedings, and 21 book chapters. He is editor/co-editor of 18 books/conference proceedings and 1 scientific book. His papers are highly cited, as proven form the total number of citations (more than 25000) and the value of the h-index (74) (source: Google Scholar). He was invited as keynote speaker in more than 30 international conferences and workshops. Since 2009 he is a member of the Administrative Committee of the IEEE Geoscience and Remote Sensing Society (GRSS). 

Dr. Bruzzone ranked first place in the Student Prize Paper Competition of the 1998 IEEE International Geoscience and Remote Sensing Symposium (IGARSS), Seattle, July 1998. Since that he was recipient of many international and national
honors and awards, including the recent IEEE GRSS 2015 Outstanding Service Award and the 2017 IEEE IGARSS Symposium Prize Paper Award. Dr. Bruzzone was a Guest Co-Editor of many Special Issues of international
journals. He is the co-founder of the IEEE International Workshop on the Analysis of Multi-Temporal Remote-Sensing Images (MultiTemp) series and is currently a member of the Permanent Steering Committee of this series
of workshops. Since 2003 he has been the Chair of the SPIE Conference on Image and Signal Processing for Remote Sensing. He has been the founder of the IEEE Geoscience and Remote Sensing Magazine for which he has been Editor-in-Chief between 2013-2017.Currently he is an Associate Editor for the IEEE Transactions on Geoscience and Remote Sensing. He has been Distinguished Speaker of the IEEE Geoscience and Remote Sensing Society
between 2012-2016.
\end{IEEEbiography}

\begin{IEEEbiographynophoto}{Biao Li}
was born in Zhejiang,
China, in 1968. He received the Ph.D. degrees in the
School of Electronic Science, National University of Defense
Technology, Changsha, China, in 1998, where he is currently a full professor. His current research
interests include signal processing and infrared image processing.
\end{IEEEbiographynophoto}






\end{document}